\documentclass[aps,preprint,nofootinbib]{revtex4}%
\usepackage{amsfonts}
\usepackage{amsmath}
\usepackage{amssymb}
\usepackage{graphicx}%
\providecommand{\U}[1]{\protect\rule{.1in}{.1in}}

\begin{document}
\title{The Hamiltonian formulation of tetrad gravity: three dimensional case}
\author{A.M. Frolov}
\affiliation{Department of Chemistry, University of Western Ontario, London, Canada}
\email{afrolov@uwo.ca}
\author{N. Kiriushcheva}
\affiliation{Department of Applied Mathematics and Department of Mathematics, University of
Western Ontario, London, Canada}
\email{nkiriush@uwo.ca}
\author{S.V. Kuzmin}
\affiliation{Faculty of Arts and Social Science, Huron University College and Department of
Applied Mathematics, University of Western Ontario, London, Canada}
\email{skuzmin@uwo.ca}
\keywords{Hamiltonian, tetrad gravity, three dimensions}\date{\today}

\pacs{11.10.Ef, 11.30.Cp}

\begin{abstract}
The Hamiltonian formulation of the tetrad gravity in any dimension higher than
two, using its first order form when tetrads and spin connections are treated
as independent variables, is discussed and the complete solution of the three
dimensional case is given. For the first time, applying the methods of
constrained dynamics, the Hamiltonian and constraints are explicitly derived
and the algebra of the Poisson brackets among all constraints is calculated.
The algebra of the Poisson brackets among first class secondary constraints
locally coincides with Lie algebra of the ISO(2,1) Poincar\'{e} group. All the
first class constraints of this formulation, according to the Dirac conjecture
and using the Castellani procedure, allow us to unambiguously derive the
generator of gauge transformations and find the gauge transformations of the
tetrads and spin connections which turn out to be the same found by Witten
without recourse to the Hamiltonian methods [\textit{Nucl. Phys. \textbf{B
311} (1988) 46}]. The gauge symmetry of the tetrad gravity generated by Lie
algebra of constraints is compared with another invariance, diffeomorphism.
Some conclusions about the Hamiltonian formulation in higher dimensions are
briefly discussed; in particular, that diffeomorphism invariance is
\textit{not derivable} as a \textit{gauge symmetry} from the Hamiltonian
formulation of tetrad gravity in any dimension when tetrads and spin
connections are used as independent variables.

PACS: 11.10.Ef, 11.30.Cp

\end{abstract}
\maketitle

\section{\bigskip Introduction}

The Hamiltonian formulation of General Relativity (GR) has a history going
back a half-century. On one hand, the Hamiltonian formulation of such a highly
non-trivial theory as GR is a good laboratory where general methods of
constrained dynamics \cite{Diracbook, Kurt, Gitmanbook} can be studied and
some subtle points that cannot be even seen in simple examples, can be found,
investigated and lead to further development of the method itself. On the
other hand, the correct Hamiltonian formulation of a theory is a prerequisite
to its successful canonical quantization. In this paper we consider the
Hamiltonian formulation of the tetrad gravity. Nowadays this is a more popular
formulation of GR, in particular, because it is used in Loop Quantum Gravity
\cite{GambiniPullin, Rovelli, Thiemann}. The accepted Hamiltonian formulation
of tetrad gravity leads to the so-called \textquotedblleft diffeomorphism
constraint\textquotedblright, or more precisely, the \textquotedblleft spatial
diffeomorphism constraint\textquotedblright\ \cite{GambiniPullin, Rovelli,
Thiemann} (though the word \textquotedblleft spatial\textquotedblright\ is
often omitted in the literature).

Recently it was demonstrated \cite{KKRV, myths, FKK} that the long-standing
problem of having \textit{only spatial} diffeomorphism in the Hamiltonian
formulation of metric GR \cite{Isham} is just a consequence of a non-canonical
change of variables. Without making such changes, the full diffeomorphism
invariance of the metric tensor is derivable from the Hamiltonian formulation
in all dimensions higher than two ($D>2$) \cite{KKRV, myths}. This result
suggests the necessity of reconsidering also the Hamiltonian formulation of
tetrad gravity, especially because the accepted Hamiltonian formulation was
performed using a change of variables for tetrads \cite{DI} similar to metric
gravity and this has led to the same \textquotedblleft diffeomorphism
constraint\textquotedblright\ which is \textit{only spatial}.

Moreover, the three dimensional case of tetrad gravity poses additional
questions. For instance, what is the \textit{gauge} symmetry of the tetrad
gravity in three dimensions? In some papers it is written that the gauge
symmetry is Poincar\'{e} symmetry \cite{Witten}, in others that it is Lorentz
symmetry plus diffeomorphism \cite{Carlip}\footnote{In some papers Lorentz
symmetry plus diffeomorphism are even called the Poincar\'{e} gauge symmetry
(see, for example, \cite{Gren, Ali}).}, or that there exists various ways to
define the constraints of tetrad gravity leading to different gauge
transformations \cite{Nelson}. According to \cite{Mats2}, two symmetries are
present and we have to decide \textquotedblleft what is a gauge symmetry and
what is not\textquotedblright. We think that this is the right question but
the answer should not depend on \textit{our} decision or desire. The
Hamiltonian method is the perfect instrument to find the unique answer to the
question what the gauge symmetry is.

To the best of our knowledge, despite the existence of numerous review
articles, living reviews, and books, e.g. \cite{Mats}, \cite{livingreview},
and \cite{Carlipbook}, there is no\textit{ complete} Hamiltonian formulation
of tetrad gravity in three dimensions. The only papers that somehow related to
tetrad gravity in three dimensions are: the work due to Blagojevi\'{c} and
Cvetkovi\'{c} \cite{Blag} where all steps of the Dirac procedure were
performed, but for the three dimensional Mielke-Baekler model; and
Blagojevi\'{c} in \cite{Blag-book} performed the Hamiltonian analysis but for
the Chern-Simons action. \textit{A complete} Hamiltonian formulation means
that \textit{all steps} of the Dirac procedure should be performed
\cite{Diracbook, Kurt, Gitmanbook}: (i) momenta are introduced to all
variables leading to the primary constraints, (ii) the Hamiltonian is found,
(iii) the time development of constraints is considered until (iv) the closure
of the Dirac procedure is reached, (v) all constraints are classified as being
first and second class, (vi) second class constraints are eliminated
(Hamiltonian reduction) \cite{AoP}, (vii) a gauge generator, according to the
Dirac conjecture \cite{Diracbook}, is constructed from all the first class
constraints using one of the available methods \cite{Cast, Hen, Ban}%
\footnote{We would like to note that the methods of \cite{Hen, Ban} should be
applied with a great caution (see \cite{Affine}).}, and (viii) this gauge
generator is used to derive unambiguously the gauge transformations of all
fields. If some of these steps are missing or implemented incorrectly then we
cannot be sure that the correct gauge symmetry has been found.

The first attempts to interpret GR as a gauge theory started from work of
Utiyama \cite{Utiyama}.\footnote{Actually, before Utiyama's attempt, Weyl
introduced what we know now as a principle of gauge invariance in his attempt
to unify electricity and gravitation \cite{Weyl}.} In his approach, in the
same way as was done in Yang-Mills theory \cite{YM}, by postulating the
invariance of a system under a certain group of transformations it is possible
to introduce a new compensating field, determine the form of interaction, and
construct the modified Lagrangian which makes the action invariant under the
given transformations. Utiyama applied his method to the gravitational field
using local Lorentz transformations for vierbein fields. Later Kibble
\cite{Kibble} extended Utiyama's prescription by considering the group of
inhomogeneous Lorentz transformations, Poincar\'{e} group, (though he switched
from the translational parameters of the Poincar\'{e} symmetry to the
parameters which \textquotedblleft specify a general coordinate
transformation\textquotedblright, e.g. diffeomorphism transformation).
Localization of Poincar\'{e} symmetry leads to the Poincar\'{e} gauge theory
of gravity (PGTG) (see, e.g. \cite{Frolov} and for the Poincar\'{e}-Weyl
theory \cite{BFZ}). All these approaches have the same feature: the aim to
construct a theory from a given gauge symmetry rather than to derive a gauge
symmetry for a given Lagrangian. This is the main disadvantage of such methods
as they cannot be used for the systems with unknown \textit{a priori} gauge invariance.

In this paper we explore another approach. We do not relate our analysis to
the Chern-Simons action (as was done by Witten in \cite{Witten}), we do not
perform any change of variables (even canonical), and do not use any
formulation which is specific to a particular dimension (like Plebanski's
\cite{Plebanski} for dimension $D=4$). Our goal is to start from the first
order action of the tetrad gravity, in which tetrads and spin connections are
treated as independent variables, follow \textit{all} steps of the Dirac
procedure, without any assumption of what the gauge symmetry should be, and
see what gauge transformations will be \textit{derived} (or what
\textquotedblleft decision\textquotedblright\ the procedure will make). The
structure of our paper follows steps (i)-(viii) of the procedure outlined above.

In the next Section we apply the Dirac procedure to the first order
formulation of tetrad GR in any dimensions ($D>2$). The first steps are
independent of the dimension\ until we reach the point where a peculiarity of
three dimensions appears. From this stage onwards, where we must consider the
elimination of second class constraints, we restrict our analysis to three
dimensional case. In Section III we perform the Hamiltonian reduction by
eliminating second class primary constraints and the corresponding pairs of
canonical variables, and then derive the explicit expressions for the
secondary first class constraints and the Hamiltonian. The closure of the
Dirac procedure and Poisson brackets (PB) among all constraints are given in
Section IV where it is demonstrated that PB algebra of secondary first class
constraints coincides with Lie algebra of the ISO(2,1) Poincar\'{e} group. In
Section V, using the Dirac conjecture and the Castellani procedure, we derive
the gauge generator from all the first class constraints and their PB algebra
which was found in the previous Section. Both gauge parameters presented in
the generator turn out to have only internal (\textquotedblleft
Lorentz\textquotedblright) indices and describe rotations and translations in
the tangent space, not a diffeomorphism. When this gauge generator acts on
fields, it gives gauge transformations which are equivalent to Witten's result
\cite{Witten} obtained for $D=3$ without the use of the Dirac procedure. This
unique gauge symmetry which has been derived for the tetrad gravity in its
first order form is compared with another non-gauge symmetry, diffeomorphism,
in Section VI. Our consideration is based on the original variables, tetrads
and spin connections, without making even a canonical change of variables and
without specialization of either the variables or the form of the original
action to any particular dimension. This allows us to use the three
dimensional case as a guide for higher dimensions and to draw some conclusions
about the Hamiltonian formulation of the tetrad gravity in higher dimensions.
This discussion is presented in Section VII. The preliminary results on the
analysis of the tetrad gravity in higher dimensions are reported in
\cite{Report, gauge, Darboux}.

\section{The Hamiltonian and constraints}

To compare our results with previous incomplete attempts of the Hamiltonian
formulation, we start our analysis from the Einstein-Cartan (EC) Lagrangian of
tetrad gravity written in its first order form (found, e.g., in
\cite{Schwinger, CNP})%

\begin{equation}
L=-e\left(  e^{\mu\left(  \alpha\right)  }e^{\nu\left(  \beta\right)  }%
-e^{\nu\left(  \alpha\right)  }e^{\mu\left(  \beta\right)  }\right)  \left(
\omega_{\nu\left(  \alpha\beta\right)  ,\mu}+\omega_{\mu\left(  \alpha
\gamma\right)  }\omega_{\nu~~\beta)}^{~(\gamma}\right)  ,\label{eqnT1}%
\end{equation}
where the covariant tetrads $e_{\gamma\left(  \rho\right)  }$ and the spin
connections $\omega_{\nu\left(  \alpha\beta\right)  }$ are treated as
independent fields, and $e=\det\left(  {e_{\gamma\left(  \rho\right)  }%
}\right)  $. We assume that the inverse $e^{\mu\left(  \alpha\right)  }$ of
the tetrad field ${e_{\gamma\left(  \rho\right)  }}$ exists and $e^{\mu\left(
\alpha\right)  }e_{\mu\left(  \beta\right)  }=\delta_{\left(  \beta\right)
}^{\left(  \alpha\right)  }$, $e^{\mu\left(  \alpha\right)  }e_{\nu\left(
\alpha\right)  }=\delta_{\nu}^{\mu}$. Indices in brackets (..) denote the
internal (\textquotedblleft Lorentz\textquotedblright) indices, whereas
indices without brackets are external or \textquotedblleft
world\textquotedblright\ indices. Internal and external indices are raised and
lowered by the Minkowski tensor $\eta_{(\alpha)(\beta)}=\left(
-,+,+,...\right)  $ and the metric tensor $g_{\mu\nu}=e_{\mu\left(
\alpha\right)  }e_{\nu}^{\left(  \alpha\right)  }$, respectively. For the
tetrad gravity, the first order form of (\ref{eqnT1}) and second order
formulations are equivalent in all dimensions, except of $D=2$. (On the
Hamiltonian formulation of tetrad gravity when $D=2$ see \cite{2DGKK}.)
Because we are interested in obtaining a formulation valid in all dimensions,
we will not specialize our notation to a particular dimension (as, e.g.,
\cite{Witten, Plebanski}), imposing only one restriction: $D>2$.

To make the analysis more transparent, we rewrite the Lagrangian using
integration by parts and introducing a few short notations:%

\begin{equation}
L=eB^{\gamma\left(  \rho\right)  \mu\left(  \alpha\right)  \nu\left(
\beta\right)  }e_{\gamma\left(  \rho\right)  ,\mu}\omega_{\nu\left(
\alpha\beta\right)  }-eA^{\mu\left(  \alpha\right)  \nu\left(  \beta\right)
}\omega_{\mu\left(  \alpha\gamma\right)  }\omega_{\nu~~\beta)}^{~(\gamma},
\label{eqnT2}%
\end{equation}
where the coefficients $A^{\mu\left(  \alpha\right)  \nu\left(  \beta\right)
}$ and $B^{\gamma\left(  \rho\right)  \mu\left(  \alpha\right)  \nu\left(
\beta\right)  }$ are%

\begin{equation}
A^{\mu\left(  \alpha\right)  \nu\left(  \beta\right)  }\equiv e^{\mu\left(
\alpha\right)  }e^{\nu\left(  \beta\right)  }-e^{\mu\left(  \beta\right)
}e^{\nu\left(  \alpha\right)  }\label{eqnT3}%
\end{equation}
and
\begin{equation}
B^{\gamma\left(  \rho\right)  \mu\left(  \alpha\right)  \nu\left(
\beta\right)  }\equiv e^{\gamma\left(  \rho\right)  }A^{\mu\left(
\alpha\right)  \nu\left(  \beta\right)  }+e^{\gamma\left(  \alpha\right)
}A^{\mu\left(  \beta\right)  \nu\left(  \rho\right)  }+e^{\gamma\left(
\beta\right)  }A^{\mu\left(  \rho\right)  \nu\left(  \alpha\right)
}.\label{eqnT4}%
\end{equation}
The symmetries of $A^{\mu\left(  \alpha\right)  \nu\left(  \beta\right)  }$
and $B^{\gamma\left(  \rho\right)  \mu\left(  \alpha\right)  \nu\left(
\beta\right)  }$ follow from their definitions: e.g., $A^{\mu\left(
\alpha\right)  \nu\left(  \beta\right)  }=A^{\nu\left(  \beta\right)
\mu\left(  \alpha\right)  }$, $A^{\mu\left(  \alpha\right)  \nu\left(
\beta\right)  }=-A^{\nu\left(  \alpha\right)  \mu\left(  \beta\right)  }$,
$A^{\mu\left(  \alpha\right)  \nu\left(  \beta\right)  }=-A^{\mu\left(
\beta\right)  \nu\left(  \alpha\right)  }$. Similar antisymmetry properties
hold for $B^{\gamma\left(  \rho\right)  \mu\left(  \alpha\right)  \nu\left(
\beta\right)  }$. In (\ref{eqnT4}) the second and third terms can be obtained
by a cyclic permutation of the internal indices $\rho\alpha\beta
\rightarrow\alpha\beta\rho\rightarrow\beta\rho\alpha$ (keeping external
indices in the same position). $B^{\gamma\left(  \rho\right)  \mu\left(
\alpha\right)  \nu\left(  \beta\right)  }$ can also be presented in different
form with cyclic permutations of external indices (keeping internal indices in
the same position)
\begin{equation}
B^{\gamma\left(  \rho\right)  \mu\left(  \alpha\right)  \nu\left(
\beta\right)  }=e^{\gamma\left(  \rho\right)  }A^{\mu\left(  \alpha\right)
\nu\left(  \beta\right)  }+e^{\mu\left(  \rho\right)  }A^{\nu\left(
\alpha\right)  \gamma\left(  \beta\right)  }+e^{\nu\left(  \rho\right)
}A^{\gamma\left(  \alpha\right)  \mu\left(  \beta\right)  }.\label{eqnT5}%
\end{equation}
These form, (\ref{eqnT4}) and (\ref{eqnT5}), are useful in calculations. As
follows from their antisymmetry, $A^{\mu\left(  \alpha\right)  \nu\left(
\beta\right)  }$ and $B^{\gamma\left(  \rho\right)  \mu\left(  \alpha\right)
\nu\left(  \beta\right)  }$ equal zero when two external or two internal
indices have the same value. The properties of $A,B$ and similar functions are
collected in Appendix A.

As in any first order formulation, the Hamiltonian analysis of first order
tetrad gravity leads to primary constraints equal in number to the number of
independent fields. Introducing momenta for all fields%

\[
\pi^{\mu\left(  \alpha\right)  }=\frac{\delta L}{\delta e_{\mu\left(
\alpha\right)  ,0}}~,\qquad\Pi^{\mu\left(  \alpha\beta\right)  }=\frac{\delta
L}{\delta\omega_{\mu\left(  \alpha\beta\right)  ,0}}~,
\]
we obtain the following set of primary constraints%

\begin{equation}
\pi^{\gamma\left(  \rho\right)  }-\frac{\delta}{\delta e_{\gamma\left(
\rho\right)  ,0}}\left(  eB^{\gamma\left(  \rho\right)  \mu\left(
\alpha\right)  \nu\left(  \beta\right)  }e_{\gamma\left(  \rho\right)  ,\mu
}\omega_{\nu\left(  \alpha\beta\right)  }\right)  =\pi^{\gamma\left(
\rho\right)  }-eB^{\gamma\left(  \rho\right)  0\left(  \alpha\right)
\nu\left(  \beta\right)  }\omega_{\nu\left(  \alpha\beta\right)  }\approx0,
\label{eqnT6}%
\end{equation}

\begin{equation}
\Pi^{\mu\left(  \alpha\beta\right)  }\approx0. \label{eqnT7}%
\end{equation}

The fundamental Poisson brackets are%

\begin{equation}
\left\{  e_{\mu\left(  \alpha\right)  }\left(  \mathbf{x}\right)  ,\pi
^{\gamma\left(  \rho\right)  }\left(  \mathbf{y}\right)  \right\}
=\delta_{\mu}^{\gamma}\delta_{\left(  \alpha\right)  }^{\left(  \rho\right)
}\delta\left(  \mathbf{x}-\mathbf{y}\right)  ,\left.  {}\right.  \left\{
\omega_{\lambda\left(  \alpha\beta\right)  }\left(  \mathbf{x}\right)
,\Pi^{\rho\left(  \mu\nu\right)  }\left(  \mathbf{y}\right)  \right\}
=\tilde{\Delta}_{\left(  \alpha\beta\right)  }^{\left(  \mu\nu\right)  }%
\delta_{\lambda}^{\rho}\delta\left(  \mathbf{x}-\mathbf{y}\right)  \label{FPB}%
\end{equation}
where%

\[
\tilde{\Delta}_{\left(  \alpha\beta\right)  }^{\left(  \mu\nu\right)  }%
\equiv\frac{1}{2}\left(  \delta_{\left(  \alpha\right)  }^{\left(  \mu\right)
}\delta_{\left(  \beta\right)  }^{\left(  \nu\right)  }-\delta_{\left(
\alpha\right)  }^{\left(  \nu\right)  }\delta_{\left(  \beta\right)
}^{\left(  \mu\right)  }\right)  .
\]
(Note that in the text we often write a PB without the factor $\delta\left(
\mathbf{x}-\mathbf{y}\right)  $).

From the antisymmetry of $B^{\gamma\left(  \rho\right)  \mu\left(
\alpha\right)  \nu\left(  \beta\right)  }$ we immediately obtain for
$\pi^{0\left(  \rho\right)  }$%

\begin{equation}
\pi^{0\left(  \rho\right)  }\approx0, \label{eqnT8}%
\end{equation}
and for $\pi^{k\left(  \rho\right)  }$%

\begin{equation}
\pi^{k\left(  \rho\right)  }-eB^{k\left(  \rho\right)  0\left(  \alpha\right)
\nu\left(  \beta\right)  }\omega_{\nu\left(  \alpha\beta\right)  }\approx0.
\label{eqnT9}%
\end{equation}
Here and everywhere below in our paper we shall apply the usual convention:
Greek letters for \textquotedblleft spacetime\textquotedblright\ (both
internal and external) indices, e.g. $\alpha=0,1,2,...,D-1$, $\beta
=0,1,2,...,D-1$, and Latin letters for \textquotedblleft
space\textquotedblright\ indices $k=1,2,...,D-1$, $m=1,2,...,D-1$, etc.

All primary constraints are now identified and the Hamiltonian density takes
the form%

\[
H=\pi^{0\left(  \rho\right)  }\dot{e}_{0\left(  \rho\right)  }+\left(
\pi^{k\left(  \rho\right)  }-eB^{k\left(  \rho\right)  0\left(  \alpha\right)
\nu\left(  \beta\right)  }\omega_{\nu\left(  \alpha\beta\right)  }\right)
\dot{e}_{k\left(  \rho\right)  }+\Pi^{\mu\left(  \alpha\beta\right)  }%
\dot{\omega}_{\mu\left(  \alpha\beta\right)  }%
\]

\begin{equation}
-eB^{\gamma\left(  \rho\right)  k\left(  \alpha\right)  \nu\left(
\beta\right)  }e_{\gamma\left(  \rho\right)  ,k}\omega_{\nu\left(  \alpha
\beta\right)  }+eA^{\mu\left(  \alpha\right)  \nu\left(  \beta\right)  }%
\omega_{\mu\left(  \alpha\gamma\right)  }\omega_{\nu~~\beta)}^{~(\gamma}.
\label{eqnT10}%
\end{equation}

There should be second class among these primary constraints because the
constraint (\ref{eqnT9}) contains connections $\omega_{\nu\left(  \alpha
\beta\right)  }$ and the PBs of at least some of them with $\Pi^{\mu\left(
\alpha\beta\right)  }$ are not zero (in particular, $\left\{  \pi^{k\left(
\rho\right)  },\Pi^{m\left(  \alpha\beta\right)  }\right\}  =-eB^{k\left(
\rho\right)  0\left(  \alpha\right)  m\left(  \beta\right)  }$). To clarify
this and to see what connections are present in (\ref{eqnT9}), we further
separate $\pi^{k\left(  \rho\right)  }$ into components (using antisymmetry of
$B^{\gamma\left(  \rho\right)  \mu\left(  \alpha\right)  \nu\left(
\beta\right)  }$)%

\begin{equation}
\pi^{k\left(  m\right)  }-2eB^{k\left(  m\right)  0\left(  q\right)  p\left(
0\right)  }\omega_{p\left(  q0\right)  }-eB^{k\left(  m\right)  0\left(
p\right)  n\left(  q\right)  }\omega_{n\left(  pq\right)  }\approx0,
\label{eqnT11}%
\end{equation}

\begin{equation}
\pi^{k\left(  0\right)  }-eB^{k\left(  0\right)  0\left(  p\right)  m\left(
q\right)  }\omega_{m\left(  pq\right)  }\approx0. \label{eqnT12}%
\end{equation}
This form shows the explicit appearance of particular connections
($\omega_{m\left(  pq\right)  }$ or $\omega_{p\left(  q0\right)  }$) in the
primary constraints. There are no connections with the \textquotedblleft
temporal\textquotedblright\ external index in (\ref{eqnT11}) and
(\ref{eqnT12}) and, correspondingly, the primary constraints $\Pi^{0\left(
\alpha\beta\right)  }$ commute with the rest of primary constraints, and
therefore, the constraints $\Pi^{0\left(  \alpha\beta\right)  }$ are first
class at this stage.

One group of constraints, which has the same form in all dimensions $D>2$,%

\begin{equation}
\pi^{k\left(  m\right)  }-2eB^{k\left(  m\right)  0\left(  q\right)  p\left(
0\right)  }\omega_{p\left(  q0\right)  }-eB^{k\left(  m\right)  0\left(
p\right)  n\left(  q\right)  }\omega_{n\left(  pq\right)  }\approx0,\left.
{}\right.  \Pi^{p\left(  q0\right)  }\approx0,\label{eqnT13}%
\end{equation}
form a second class subset, and using them one pair of canonical variables
($\omega_{p\left(  q0\right)  },\Pi^{p\left(  q0\right)  }$) can be now
eliminated. These constraints, (\ref{eqnT13}), are not of a special form
\cite{Gitmanbook}, but they are linear in $\omega_{p\left(  q0\right)  }$ and
$\Pi^{p\left(  q0\right)  }$ and the coefficient in front of $\omega_{p\left(
q0\right)  }$ in (\ref{eqnT13}) does not depend either on $\omega_{p\left(
q0\right)  }$ or  $\Pi^{p\left(  q0\right)  }$, so after their elimination,
the PBs among the remaining canonical variables will not change (i.e., they
are the same as the Dirac brackets).

To eliminate this pair, ($\omega_{p\left(  q0\right)  },\Pi^{p\left(
q0\right)  }$), we have to solve (\ref{eqnT11}) for $\omega_{p\left(
q0\right)  }$ in terms of $\omega_{n\left(  pq\right)  }$ and $\pi^{k\left(
m\right)  }$, and substitute this solution, as well as $\Pi^{p\left(
q0\right)  }=0$, into the total Hamiltonian. The solution to equation
(\ref{eqnT11}) for $\omega_{p\left(  q0\right)  }$ exists in all dimensions
$D>2$. In fact, it becomes especially simple if one notices that%

\begin{equation}
B^{k\left(  m\right)  0\left(  q\right)  p\left(  0\right)  }=-e^{0\left(
0\right)  }E^{k\left(  m\right)  p\left(  q\right)  } \label{eqnT14}%
\end{equation}
where%

\begin{equation}
E^{k\left(  m\right)  p\left(  q\right)  }\equiv\gamma^{k\left(  m\right)
}\gamma^{p\left(  q\right)  }-\gamma^{k\left(  q\right)  }\gamma^{p\left(
m\right)  } \label{eqnT15}%
\end{equation}
and%

\begin{equation}
\gamma^{k\left(  m\right)  }\equiv e^{k\left(  m\right)  }-\frac{e^{k\left(
0\right)  }e^{0\left(  m\right)  }}{e^{0\left(  0\right)  }} \label{eqnT16}%
\end{equation}
with properties%

\begin{equation}
\gamma^{m\left(  p\right)  }e_{m\left(  q\right)  }=\delta_{\left(  q\right)
}^{\left(  p\right)  },\left.  {}\right.  \left.  {}\right.  \gamma^{n\left(
q\right)  }e_{m\left(  q\right)  }=\delta_{m}^{n}.\label{eqnT17}%
\end{equation}
$E^{k\left(  m\right)  p\left(  q\right)  }$ is also antisymmetric (i.e.
$E^{k\left(  m\right)  p\left(  q\right)  }=-E^{p\left(  m\right)  k\left(
q\right)  }=-E^{k\left(  q\right)  p\left(  m\right)  }$) and equals to zero
if $k=p$ or $\left(  m\right)  =\left(  q\right)  $.

For any dimension ($D>2$) we can define the inverse of $E^{k\left(  m\right)
p\left(  q\right)  }$%

\begin{equation}
I_{m\left(  q\right)  a\left(  b\right)  }\equiv\frac{1}{D-2}e_{m\left(
q\right)  }e_{a\left(  b\right)  }-e_{m\left(  b\right)  }e_{a\left(
q\right)  }. \label{eqnT18}%
\end{equation}
It is easy to check that
\begin{equation}
I_{m\left(  q\right)  a\left(  b\right)  }E^{a\left(  b\right)  n\left(
p\right)  }=E^{n\left(  p\right)  a\left(  b\right)  }I_{a\left(  b\right)
m\left(  q\right)  }=\delta_{m}^{n}\delta_{\left(  q\right)  }^{\left(
p\right)  }. \label{eqnT19}%
\end{equation}

Using the above notation, the solution of (\ref{eqnT11}) can be written in the form%

\begin{equation}
\omega_{k\left(  q0\right)  }=-\frac{1}{2ee^{0\left(  0\right)  }}I_{k\left(
q\right)  m\left(  p\right)  }\pi^{m\left(  p\right)  }+\frac{1}{2e^{0\left(
0\right)  }}I_{k\left(  q\right)  m\left(  p\right)  }B^{m\left(  p\right)
0\left(  a\right)  n\left(  b\right)  }\omega_{n\left(  ab\right)  }.
\label{eqnT20}%
\end{equation}

Hence we see that the constraint (\ref{eqnT11}) can be solved for
$\omega_{p\left(  q0\right)  }$ in any dimension $D>2$, because of the
existence of the inverse $I_{k\left(  q\right)  m\left(  p\right)  }$ and
because of the same number of equations and unknowns in (\ref{eqnT11}), i.e.
$\left[  \pi^{m\left(  p\right)  }\right]  =\left[  \omega_{k\left(
q0\right)  }\right]  =\left(  D-1\right)  ^{2}$ (where $\left[  X\right]  $
indicates the number of components of a field $X$).

The second constraint, (\ref{eqnT12}), cannot be solved unambiguously for
$\omega_{k\left(  pq\right)  }$ because the number of equations, $\left[
\pi^{k\left(  0\right)  }\right]  =D-1$, and the number of unknowns, $\left[
\omega_{m\left(  pq\right)  }\right]  =\frac{1}{2}\left(  D-1\right)
^{2\text{ }}\left(  D-2\right)  $, are, in general, different. This difference
depends on the dimension of spacetime:%

\begin{equation}
\left[  \omega_{m\left(  pq\right)  }\right]  -\left[  \pi^{k\left(  0\right)
}\right]  =\frac{1}{2}\left(  D-1\right)  ^{2\text{ }}\left(  D-2\right)
-\left(  D-1\right)  =\frac{1}{2}\left(  D-1\right)  D\left(  D-3\right)  .
\label{eqnT21}%
\end{equation}

In dimensions $D>3$ we can choose only a subset of these variables for
elimination which is not unique and, more importantly, this procedure will
destroy the covariant form of constraints. It also creates difficulties in a
consistent elimination of these fields. The components of momenta (primary
constraints) that would be left after this elimination presumably would lead
to secondary constraints that could be eliminated at the next stage of the
Hamiltonian reduction (solving this problem in different order or mixing a
primary second class pair with pairs of primary and secondary constraints is a
difficult task). The detail of this analysis for $D>3$ can be found in
\cite{Report, Darboux}.

When $D=3$, the difference in (\ref{eqnT21}) is zero. We have $\left[
\pi^{k\left(  0\right)  }\right]  =\left[  \omega_{m\left(  pq\right)
}\right]  =2$, or two equations in two unknowns in (\ref{eqnT12}). This
drastically simplifies calculations. What is important, we have unambiguously
one more pair of second class primary constraints, and all connections with
\textquotedblleft spatial\textquotedblright\ external indices and their
conjugate momenta are eliminated at this stage leading immediately to the
Hamiltonian and primary constraints which have vanishing PBs. In next Section
we analyze this case ($D=3$).

\section{The Hamiltonian analysis of the tetrad gravity in D=3}

From this point onwards, we specialize to the case of $D=3$. The same number
of equations and unknowns in (\ref{eqnT11}) and (\ref{eqnT12}) allows us to
eliminate all connections with \textquotedblleft spatial\textquotedblright%
\ external indices by solving the primary second class constraints. Moreover,
there are additional simplifications that occur only for $D=3$. First of all,
in this dimension the constraint (\ref{eqnT11}) becomes simpler because the
second term is zero (there are three \textquotedblleft
spatial\textquotedblright\ internal indices in $B^{\gamma\left(  m\right)
\mu\left(  p\right)  \nu\left(  q\right)  }$ and in when $D=3$ at least two of
them have to be equal which, based on the properties of $B^{\gamma\left(
\rho\right)  \mu\left(  \alpha\right)  \nu\left(  \beta\right)  }$, gives
$B^{k\left(  m\right)  0\left(  p\right)  n\left(  q\right)  }=0$). This leads
also to a separation of the components of the spin connections among the
primary constrains. Equations (\ref{eqnT11}) and (\ref{eqnT12}) become%

\begin{equation}
\pi^{k\left(  m\right)  }-2eB^{k\left(  m\right)  0\left(  q\right)  p\left(
0\right)  }\omega_{p\left(  q0\right)  }\approx0, \label{eqnT22}%
\end{equation}

\begin{equation}
\pi^{k\left(  0\right)  }-eB^{k\left(  0\right)  0\left(  p\right)  m\left(
q\right)  }\omega_{m\left(  pq\right)  }\approx0. \label{eqnT23}%
\end{equation}

The Hamiltonian in this case is%

\[
H=\pi^{0\left(  \rho\right)  }\dot{e}_{0\left(  \rho\right)  }+\left(
\pi^{k\left(  0\right)  }-eB^{k\left(  0\right)  0\left(  p\right)  m\left(
q\right)  }\omega_{m\left(  pq\right)  }\right)  \dot{e}_{k\left(  0\right)
}+\left(  \pi^{k\left(  m\right)  }-2eB^{k\left(  m\right)  0\left(  q\right)
p\left(  0\right)  }\omega_{p\left(  q0\right)  }\right)  \dot{e}_{k\left(
m\right)  }%
\]

\begin{equation}
+\Pi^{m\left(  \alpha\beta\right)  }\dot{\omega}_{m\left(  \alpha\beta\right)
}+\Pi^{0\left(  \alpha\beta\right)  }\dot{\omega}_{0\left(  \alpha
\beta\right)  }-eB^{\gamma\left(  \rho\right)  k\left(  \alpha\right)
\nu\left(  \beta\right)  }e_{\gamma\left(  \rho\right)  ,k}\omega_{\nu\left(
\alpha\beta\right)  }+eA^{\mu\left(  \alpha\right)  \nu\left(  \beta\right)
}\omega_{\mu\left(  \alpha\gamma\right)  }\omega_{\nu~~\beta)}^{~(\gamma}.
\label{eqnT24}%
\end{equation}
One group of constraints allows us to perform the Hamiltonian reduction%

\begin{equation}
\Pi^{m\left(  p0\right)  }=0, \label{eqnT25}%
\end{equation}

\begin{equation}
\omega_{k\left(  q0\right)  }=-\frac{1}{2ee^{0\left(  0\right)  }}I_{k\left(
q\right)  m\left(  p\right)  }\pi^{m\left(  p\right)  }. \label{eqnT26}%
\end{equation}
Similarly, for the second group of constraints we have%

\begin{equation}
\Pi^{m\left(  pq\right)  }=0 \label{eqnT27}%
\end{equation}
and we need to solve (\ref{eqnT23}) for $\omega_{m\left(  pq\right)  }$. Using
the symmetries of $B^{\gamma\left(  \rho\right)  \mu\left(  \alpha\right)
\nu\left(  \beta\right)  }$ and (\ref{eqnT14}) we can rewrite (\ref{eqnT23})
in the form%

\begin{equation}
\pi^{k\left(  0\right)  }+ee^{0\left(  0\right)  }E^{k\left(  q\right)
m\left(  p\right)  }\omega_{m\left(  qp\right)  }=0. \label{eqnT28}%
\end{equation}

When $D=3$ there are only two independent components of $\omega_{m\left(
pq\right)  }$: $\omega_{1\left(  12\right)  }$ and $\omega_{2\left(
12\right)  }$. Writing explicitly (\ref{eqnT28}) in components and using the
antisymmetry of $E^{k\left(  q\right)  m\left(  p\right)  }$, the solution of
this equation can be found and presented in a short, manifestly
\textquotedblleft covariant\textquotedblright, form%

\begin{equation}
\omega_{m\left(  pq\right)  }=-\frac{1}{2ee^{0\left(  0\right)  }}I_{m\left(
p\right)  k\left(  q\right)  }\pi^{k\left(  0\right)  }.\label{eqnT29}%
\end{equation}
Note that (\ref{eqnT29}) is the result of peculiarities of the three
dimensional case, contrary to (\ref{eqnT11}) which is valid in any dimension
$D>2$.

Substitution of (\ref{eqnT25}), (\ref{eqnT26}) and (\ref{eqnT27}),
(\ref{eqnT29}) into (\ref{eqnT10}) gives us the reduced Hamiltonian with a
fewer number of canonical variables and primary constraints%

\begin{equation}
H=\pi^{0\left(  \rho\right)  }\dot{e}_{0\left(  \rho\right)  }+\Pi^{0\left(
\alpha\beta\right)  }\dot{\omega}_{0\left(  \alpha\beta\right)  }-L_{\left(
\omega_{m\left(  pq\right)  }=\omega_{m\left(  pq\right)  }\left(
\pi^{k\left(  0\right)  }\right)  ,~\omega_{k\left(  q0\right)  }%
=\omega_{k\left(  q0\right)  }\left(  \pi^{m\left(  p\right)  }\right)
\right)  } \label{eqnT30}%
\end{equation}
where we have explicitly separated terms proportional to $\omega_{0\left(
\alpha\beta\right)  }$ in the canonical Hamiltonian%

\begin{align*}
H_{c}  &  =-L\left(  \text{without \textquotedblleft
velocities\textquotedblright}\right)  =-2eB^{\gamma\left(  \rho\right)
k\left(  p\right)  m\left(  0\right)  }e_{\gamma\left(  \rho\right)  ,k}%
\omega_{m\left(  p0\right)  }\\
&  -eB^{\gamma\left(  \rho\right)  k\left(  p\right)  m\left(  q\right)
}e_{\gamma\left(  \rho\right)  ,k}\omega_{m\left(  pq\right)  }+eA^{n\left(
p\right)  m\left(  q\right)  }\omega_{n\left(  p0\right)  }\omega
_{m~~q)}^{~(0}+2eA^{n\left(  0\right)  m\left(  q\right)  }\omega_{n\left(
0r\right)  }\omega_{m~~q)}^{~(r}%
\end{align*}

\begin{equation}
+\left(  2eA^{0\left(  \alpha\right)  m\left(  \gamma\right)  }\omega
_{m~~\gamma)}^{~(\beta}-eB^{\gamma\left(  \rho\right)  k\left(  \alpha\right)
0\left(  \beta\right)  }e_{\gamma\left(  \rho\right)  ,k}\right)
\omega_{0\left(  \alpha\beta\right)  }. \label{eqnT31}%
\end{equation}
\qquad

Note that in (\ref{eqnT31}) there are no terms quadratic in connections with
\textquotedblleft spatial\textquotedblright\ external indices when $D=3$. This
is the result of antisymmetry of $A^{\mu\left(  \alpha\right)  \nu\left(
\beta\right)  }$ and spin connections $\omega_{\gamma\left(  \alpha
\beta\right)  }$ in $\alpha$ and $\beta$, and since when $D=3$ the
\textquotedblleft spatial\textquotedblright\ index can take only two values, 1
and 2.

The reduced total Hamiltonian is%

\begin{equation}
H_{T}=\pi^{0\left(  \rho\right)  }\dot{e}_{0\left(  \rho\right)  }%
+\Pi^{0\left(  \alpha\beta\right)  }\dot{\omega}_{0\left(  \alpha\beta\right)
}+H_{c}\left(  e_{\mu\left(  \alpha\right)  },\pi^{m\left(  \alpha\right)
},\omega_{0\left(  \alpha\beta\right)  }\right)  \label{eqnT32}%
\end{equation}
where, after a few simple rearrangements, the canonical Hamiltonian when $D=3$ becomes%

\begin{align}
H_{c}  &  =e_{0\left(  \rho\right)  ,k}\pi^{k\left(  \rho\right)
}-e_{0\left(  \rho\right)  }\frac{1}{4ee^{0\left(  0\right)  }}I_{m\left(
q\right)  n\left(  r\right)  }\pi^{m\left(  q\right)  }\left(  \eta^{\left(
\rho\right)  \left(  0\right)  }\pi^{n\left(  r\right)  }-2\eta^{\left(
\rho\right)  \left(  r\right)  }\pi^{n\left(  0\right)  }\right)
\label{eqnT32a}\\
&  -\frac{1}{2}\left(  \pi^{k\left(  \alpha\right)  }e_{k}^{\left(
\beta\right)  }-\pi^{k\left(  \beta\right)  }e_{k}^{\left(  \alpha\right)
}+2eB^{\gamma\left(  \rho\right)  k\left(  \alpha\right)  0\left(
\beta\right)  }e_{\gamma\left(  \rho\right)  ,k}\right)  \omega_{0\left(
\alpha\beta\right)  }.\nonumber
\end{align}

To summarize, after the reduction, we have the Hamiltonian with simple primary
constraints $\pi^{0\left(  \rho\right)  }$ and $\Pi^{0\left(  \alpha
\beta\right)  }$ and all PBs among them are zero (as they are just
fundamental, canonical, variables of this formulation). With such a simple
Hamiltonian and a trivial PB algebra of primary constraints, the secondary
constraints follow immediately from conservation of primary constraints%

\begin{equation}
\dot{\pi}^{0\left(  \rho\right)  }=\left\{  \pi^{0\left(  \rho\right)  }%
,H_{T}\right\}  =-\frac{\delta H_{c}}{\delta e_{0\left(  \rho\right)  }}%
\equiv\chi^{0\left(  \rho\right)  }, \label{eqn33}%
\end{equation}

\begin{equation}
\dot{\Pi}^{0\left(  \alpha\beta\right)  }=\left\{  \Pi^{0\left(  \alpha
\beta\right)  },H_{T}\right\}  =-\frac{\delta H_{c}}{\delta\omega_{0\left(
\alpha\beta\right)  }}\equiv\chi^{0\left(  \alpha\beta\right)  }.
\label{eqnT34}%
\end{equation}
The explicit expressions for $\chi^{0\left(  \rho\right)  }$ in (\ref{eqn33}) is%

\begin{equation}
\chi^{0\left(  \rho\right)  }=\pi_{,k}^{k\left(  \rho\right)  }+\frac
{1}{4ee^{0\left(  0\right)  }}I_{m\left(  q\right)  n\left(  r\right)  }%
\pi^{m\left(  q\right)  }\left(  \eta^{\left(  \rho\right)  \left(  0\right)
}\pi^{n\left(  r\right)  }-2\eta^{\left(  \rho\right)  \left(  r\right)  }%
\pi^{n\left(  0\right)  }\right)  \label{eqnT35}%
\end{equation}
(note that, because of (\ref{eqnT29}), the form of $\chi^{0\left(
\rho\right)  }$ is also specific for $D=3$ only).

The last constraint, (\ref{eqnT34}), is obviously%

\begin{equation}
\chi^{0\left(  \alpha\beta\right)  }=\frac{1}{2}\pi^{k\left(  \alpha\right)
}e_{k}^{\left(  \beta\right)  }-\frac{1}{2}\pi^{k\left(  \beta\right)  }%
e_{k}^{\left(  \alpha\right)  }+eB^{n\left(  \rho\right)  k\left(
\alpha\right)  0\left(  \beta\right)  }e_{n\left(  \rho\right)  ,k}%
.\label{eqnT36}%
\end{equation}
We will call (\ref{eqnT35}) and (\ref{eqnT36}) the \textquotedblleft
translational\textquotedblright\ and \textquotedblleft
rotational\textquotedblright\ constraints, respectively, for reasons that will
become clear at the end of the analysis.

\section{Closure of the Dirac procedure}

To prove that the Dirac procedure closes, we have to find the time development
of secondary constraints, and check whether they produce new constraints or
not. If tertiary constraints arise, we have to continue the procedure until no
new constraints appear. If the PBs of secondary constraints with the total
Hamiltonian are zero or proportional to constraints already present, then the
procedure stops \cite{Diracbook}. The time development of the first class
constraints and the PBs amongst them and with $H_{T}$ are sufficient to find
the gauge transformations of all canonical variables \cite{Cast}.

We first compute the PBs amongst the constraints. The primary constraints
$\pi^{0\left(  \rho\right)  }$ and $\Pi^{0\left(  \alpha\beta\right)  }$ have
vanishing PBs amongst themselves%

\begin{equation}
\left\{  \pi^{0\left(  \rho\right)  },\Pi^{0\left(  \alpha\beta\right)
}\right\}  =0. \label{eqnT37b}%
\end{equation}
The rotational constraint has obviously a zero PB with the primary constraint
that generates it%

\begin{equation}
\left\{  \Pi^{0\left(  \mu\nu\right)  },\chi^{0\left(  \alpha\beta\right)
}\right\}  =0. \label{eqnT37a}%
\end{equation}
The PB of this constraint, $\chi^{0\left(  \alpha\beta\right)  }$, with the
second primary constraint is also zero%

\begin{equation}
\left\{  \pi^{0\left(  \rho\right)  },\chi^{0\left(  \alpha\beta\right)
}\right\}  =0. \label{eqnT37}%
\end{equation}
With this result it is obvious that the only contribution to the secondary
translational constraint comes from variation of that part of the Hamiltonian
(\ref{eqnT32a}) which is not proportional to the spin connections with a
\textquotedblleft temporal\textquotedblright\ external index. Because there
are no contributions proportional to the connection $\omega_{0\left(  \mu
\nu\right)  }$ in the secondary translational constraints (\ref{eqnT35}), its
PB with the primary rotational constraint is zero%

\begin{equation}
\left\{  \chi^{0\left(  \alpha\right)  },\Pi^{0\left(  \mu\nu\right)
}\right\}  =0.\label{eqnT38}%
\end{equation}
The PB among the secondary and primary translational constraints has to be
calculated. The result is%

\begin{equation}
\left\{  \pi^{0\left(  \rho\right)  },\chi^{0\left(  \alpha\right)  }\right\}
=0. \label{eqnT39}%
\end{equation}
These vanishing PBs among all primary and secondary constraints simplify the
analysis. We almost immediately can express the canonical Hamiltonian as a
linear combination of secondary constraints plus a total spatial derivative%

\begin{equation}
H_{c}=-e_{0\left(  \rho\right)  }\chi^{0\left(  \rho\right)  }-\omega
_{0\left(  \alpha\beta\right)  }\chi^{0\left(  \alpha\beta\right)  }+\left(
e_{0\left(  \rho\right)  }\pi^{k\left(  \rho\right)  }\right)  _{,k}%
.\label{eqnT40}%
\end{equation}
Taking into account the PBs among primary and secondary constraints,
constraints (\ref{eqn33}) and (\ref{eqnT34}) follow from the variation of
$H_{c}$.

Calculation of PBs among secondary constraints is straightforward though
tedious and the presence of derivatives of $e_{n\left(  \rho\right)  }$
requires the use of test functions (e.g., see \cite{testKM}). We obtain%

\begin{equation}
\left\{  \chi^{0\left(  \rho\right)  },\chi^{0\left(  \gamma\right)
}\right\}  =0, \label{qnT41}%
\end{equation}

\begin{equation}
\left\{  \chi^{0\left(  \alpha\beta\right)  },\chi^{0\left(  \rho\right)
}\right\}  =\frac{1}{2}\eta^{\left(  \beta\right)  \left(  \rho\right)  }%
\chi^{0\left(  \alpha\right)  }-\frac{1}{2}\eta^{\left(  \alpha\right)
\left(  \rho\right)  }\chi^{0\left(  \beta\right)  }, \label{eqnT42}%
\end{equation}

\begin{equation}
\left\{  \chi^{0\left(  \alpha\beta\right)  },\chi^{0\left(  \mu\nu\right)
}\right\}  =\frac{1}{2}\eta^{\left(  \beta\right)  \left(  \mu\right)  }%
\chi^{0\left(  \alpha\nu\right)  }-\frac{1}{2}\eta^{\left(  \alpha\right)
\left(  \mu\right)  }\chi^{0\left(  \beta\nu\right)  }+\frac{1}{2}%
\eta^{\left(  \beta\right)  \left(  \nu\right)  }\chi^{0\left(  \mu
\alpha\right)  }-\frac{1}{2}\eta^{\left(  \alpha\right)  \left(  \nu\right)
}\chi^{0\left(  \mu\beta\right)  }. \label{eqnT43}%
\end{equation}

Note, in calculations of (\ref{qnT41}) and (\ref{eqnT42}) we also used the
fact that the form of $\chi^{0\left(  \rho\right)  }$, (\ref{eqnT29}), is
peculiar to $D=3$. However, the PBs of $\chi^{0\left(  \alpha\beta\right)  }$
among themselves, (\ref{eqnT43}), and with primary constraints, (\ref{eqnT37a}%
) and (\ref{eqnT37}), are found without reference to $D=3$, and, as was shown
in \cite{Report}, these PBs remain valid in all dimensions ($D>2$). In papers
on group theory the brackets (\ref{eqnT42}) and (\ref{eqnT43}) usually appear
without $\frac{1}{2}$. It is easy to remove this factor if we replace
$\chi^{0\left(  \alpha\beta\right)  }$ by $2\chi^{0\left(  \alpha\beta\right)
}$. However, we do not make this replacement here, because when deriving gauge
transformations using the method of \cite{Cast} it is important to find out
what secondary constraint is produced exactly by the time development of the
corresponding primary constraint, (\ref{eqnT36}).

It is very simple to calculate time development of the secondary constraints
$\chi^{0\left(  \rho\right)  }$ and $\chi^{0\left(  \alpha\beta\right)  }$,
because $H_{c}$ is proportional to these constraints (\ref{eqnT40}), and we
have only simple local PB (there are no derivatives of $\delta$-functions
among them and this allows us to use the associative properties of the PB).
The result is:%

\begin{equation}
\dot{\chi}^{0\left(  \gamma\right)  }=\left\{  \chi^{0\left(  \gamma\right)
},H_{c}\right\}  =\frac{1}{2}\omega_{0\left(  \alpha\beta\right)  }\left(
\eta^{\left(  \beta\right)  \left(  \gamma\right)  }\chi^{0\left(
\alpha\right)  }-\eta^{\left(  \alpha\right)  \left(  \gamma\right)  }%
\chi^{0\left(  \beta\right)  }\right)  , \label{eqnT43a}%
\end{equation}

\begin{equation}
\dot{\chi}^{0\left(  \mu\nu\right)  }=\left\{  \chi^{0\left(  \mu\nu\right)
},H_{c}\right\}  =-\frac{1}{2}e_{0\left(  \rho\right)  }\left(  \eta^{\left(
\nu\right)  \left(  \rho\right)  }\chi^{0\left(  \mu\right)  }-\eta^{\left(
\mu\right)  \left(  \rho\right)  }\chi^{0\left(  \nu\right)  }\right)
\label{eqnT43c}%
\end{equation}

\[
-\frac{1}{2}\omega_{0\left(  \alpha\beta\right)  }\left(  \eta^{\left(
\alpha\right)  \left(  \mu\right)  }\chi^{0\left(  \beta\nu\right)  }%
-\eta^{\left(  \beta\right)  \left(  \mu\right)  }\chi^{0\left(  \alpha
\nu\right)  }+\eta^{\left(  \alpha\right)  \left(  \nu\right)  }\chi^{0\left(
\mu\beta\right)  }-\eta^{\left(  \beta\right)  \left(  \nu\right)  }%
\chi^{0\left(  \mu\alpha\right)  }\right)  .
\]

The above relations, (\ref{eqnT43a}) and (\ref{eqnT43c}), show that no new
constraints appear. This completes the proof that the Dirac procedure is
closed. All constraints ($\pi^{0\left(  \rho\right)  },\Pi^{0\left(
\alpha\beta\right)  },\chi^{0\left(  \rho\right)  },\chi^{0\left(  \alpha
\beta\right)  }$) are first class after elimination of $\omega_{k\left(
pq\right)  }$ and $\omega_{p\left(  q0\right)  }$ and, moreover, the PBs of
the secondary constraints, (\ref{qnT41})-(\ref{eqnT43}), form an ISO(2,1)
Poincar\'{e} algebra. This is a well defined Lie algebra: there are only
structure constants, no non-local PBs, and it is closed \textquotedblleft
off-shell\textquotedblright. (In this respect, it is similar to the gauge
invariance of Yang-Mills theory.) The same result was obtained by Witten
\cite{Witten}, though he did not use the Dirac procedure, instead he
constructed the theory starting from the Poincar\'{e} algebra.

Now let us evaluate the degree of freedom ($DOF$) in the case of $3D$ tetrad
gravity after eliminating $\omega_{k\left(  pq\right)  }$ and $\omega
_{p\left(  q0\right)  }$. Using $DOF=\#(fields)-\#(FC~~constraints)$ we obtain%

\begin{equation}
DOF=\left[  e_{\mu\left(  \rho\right)  }\right]  +\left[  \omega_{0\left(
\alpha\beta\right)  }\right]  -\left(  \left[  \pi^{0\left(  \rho\right)
}\right]  +\left[  \Pi^{0\left(  \alpha\beta\right)  }\right]  +\left[
\chi^{0\left(  \rho\right)  }\right]  +\left[  \chi^{0\left(  \alpha
\beta\right)  }\right]  \right)  =0\label{eqnT44}%
\end{equation}
as expected for the $D=3$ case.

In the literature there are some discussions about the Hamiltonian formulation
of pure three dimensional tetrad gravity, but, in fact, a complete Hamiltonian
analysis has never before been performed. In particular, in \cite{Mats2} the
Poincar\'{e} algebra is given but there is no explicit form of either the
constraints or the Hamiltonian. In \cite{Witten} the analysis was done by
comparing the three dimensional tetrad gravity with Chern-Simons theory, the
gauge transformations were given but without a derivation. The Hamiltonian
analysis of the Chern-Simons action was done in \cite{Blag-book}. We have not
found any work (including reviews and books which are dedicated to $D=3$ case,
e.g. \cite{Carlipbook}) where the gauge transformations of the three
dimensional tetrad gravity were \textit{derived} from the first class
constraints according to one of the known procedure \cite{Cast, Hen, Ban} (see
footnote 2). Such a derivation is the subject of next Section.

\section{Gauge transformations from the Castellani procedure}

We will derive the gauge transformations arising from the first class
constraints by using the Castellani procedure. This procedure \cite{Cast} is
based on a derivation of gauge generators which are defined by \textit{chains}
of first class constraints. One starts with primary first class constraint(s),
$i=1,2,...$, and construct the chain(s) $\varepsilon_{i}^{\left(  n\right)
}G_{\left(  n\right)  }^{i}$ where the $\varepsilon_{i}^{\left(  n\right)  }$
are time derivatives of order $\left(  n\right)  $\ of $i$th gauge parameter
(the maximum value of $\left(  n\right)  $ is fixed by the length of the
chain). In the case under consideration (the three dimensional tetrad gravity)
with only primary and secondary constraints $n=0$, $1$. The number of
independent gauge parameters is equal to the number of first class primary
constraints. Note that chains are unambiguously constructed once the primary
first class constraints are determined.

For tetrad gravity, we have two chains of constraints starting from the
translational ($\pi^{0\left(  \rho\right)  }$) and rotational ($\Pi^{0\left(
\alpha\beta\right)  }$) primary first class constraints. According to the
Castellani procedure, the generator is given by%

\begin{equation}
G=G_{\left(  1\right)  }^{\left(  \rho\right)  }\dot{t}_{\left(  \rho\right)
}+G_{\left(  0\right)  }^{\left(  \rho\right)  }t_{\left(  \rho\right)
}+G_{\left(  1\right)  }^{\left(  \alpha\beta\right)  }\dot{r}_{\left(
\alpha\beta\right)  }+G_{\left(  0\right)  }^{\left(  \alpha\beta\right)
}r_{\left(  \alpha\beta\right)  }.\label{eqnT45}%
\end{equation}
Here $t_{\left(  \rho\right)  }$ and $r_{\left(  \alpha\beta\right)  }$ are
gauge parameters which, as we will show later, parametrize the translational
and rotational gauge symmetries, respectively. Note that these gauge
parameters have internal indices only. It is clear even now that from the
first class constraints of tetrad gravity it is impossible to derive a
generator of the diffeomorphism invariance for tetrads and spin
connections.\footnote{Diffeomorphism was also not derivable from the
constraint structure of the Chern-Simons action \cite{Blag-book}.} The
diffeomorphism gauge parameter $\xi_{\mu}$ is of a very different nature. It
is a \textquotedblleft world\textquotedblright\ vector because it has an
external index, whereas $t_{\left(  \rho\right)  }$ and $r_{\left(
\alpha\beta\right)  }$ are \textquotedblleft world\textquotedblright\ scalars.
We will discuss the relation between the \textit{gauge} symmetry of tetrad
gravity and diffeomorphism invariance in the next Section.

The functions $G_{\left(  1\right)  }$ in (\ref{eqnT45}) are the primary constraints%

\begin{equation}
G_{\left(  1\right)  }^{\left(  \rho\right)  }=\pi^{0\left(  \rho\right)
},\left.  {}\right.  G_{\left(  1\right)  }^{\left(  \alpha\beta\right)  }%
=\Pi^{0\left(  \alpha\beta\right)  } \label{eqnT46}%
\end{equation}
and $G_{\left(  0\right)  }$ are defined using the following relations
\cite{Cast}%

\begin{equation}
G_{\left(  0\right)  }^{\left(  \rho\right)  }\left(  x\right)  =-\left\{
\pi^{0\left(  \rho\right)  }\left(  x\right)  ,H_{T}\right\}  +\int\left[
\alpha_{\left(  \gamma\right)  }^{\left(  \rho\right)  }\left(  x,y\right)
\pi^{0\left(  \gamma\right)  }\left(  y\right)  +\alpha_{\left(  \alpha
\beta\right)  }^{\left(  \rho\right)  }\left(  x,y\right)  \Pi^{0\left(
\alpha\beta\right)  }\left(  y\right)  \right]  d^{2}y, \label{eqnT47a}%
\end{equation}

\begin{equation}
G_{\left(  0\right)  }^{\left(  \alpha\beta\right)  }\left(  x\right)
=-\left\{  \Pi^{0\left(  \alpha\beta\right)  }\left(  x\right)  ,H_{T}%
\right\}  +\int\left[  \alpha_{\left(  \gamma\right)  }^{\left(  \alpha
\beta\right)  }\left(  x,y\right)  \pi^{0\left(  \gamma\right)  }\left(
y\right)  +\alpha_{\left(  \nu\mu\right)  }^{\left(  \alpha\beta\right)
}\left(  x,y\right)  \Pi^{0\left(  \nu\mu\right)  }\left(  y\right)  \right]
d^{2}y, \label{eqnT47}%
\end{equation}
where the functions $\alpha_{\left(  ..\right)  }^{\left(  ..\right)  }\left(
x,y\right)  $ have to be chosen in such a way that the chains end at primary constraints%

\begin{equation}
\left\{  G_{\left(  0\right)  }^{\sigma},H_{T}\right\}  =primary.
\label{eqnT48}%
\end{equation}

To construct the generator (\ref{eqnT45}), we have to find $\alpha_{\left(
..\right)  }^{\left(  ..\right)  }\left(  x,y\right)  $ using condition
(\ref{eqnT48}). This calculation, because of the simple PBs among the
constraints when $D=3$, is straightforward:%

\begin{equation}
\left\{  G_{\left(  0\right)  }^{\left(  \rho\right)  }\left(  x\right)
,H_{T}\right\}  =\left\{  -\chi^{0\left(  \rho\right)  }\left(  x\right)
+\int\left[  \alpha_{\left(  \gamma\right)  }^{\left(  \rho\right)  }\left(
x,y\right)  \pi^{0\left(  \gamma\right)  }\left(  y\right)  +\alpha_{\left(
\alpha\beta\right)  }^{\left(  \rho\right)  }\left(  x,y\right)  \Pi^{0\left(
\alpha\beta\right)  }\left(  y\right)  \right]  d^{2}y,H_{T}\right\}  =0,
\label{eqnT49}%
\end{equation}

\begin{equation}
\left\{  G_{\left(  0\right)  }^{\left(  \alpha\beta\right)  }\left(
x\right)  ,H_{T}\right\}  =\left\{  -\chi^{0\left(  \alpha\beta\right)
}\left(  x\right)  +\int\left[  \alpha_{\left(  \gamma\right)  }^{\left(
\alpha\beta\right)  }\left(  x,y\right)  \pi^{0\left(  \gamma\right)  }\left(
y\right)  +\alpha_{\left(  \nu\mu\right)  }^{\left(  \alpha\beta\right)
}\left(  x,y\right)  \Pi^{0\left(  \nu\mu\right)  }\left(  y\right)  \right]
d^{2}y,H_{T}\right\}  =0 \label{eqnT50}%
\end{equation}
where $H_{T}$ can be replaced by $H_{c}=-e_{0\left(  \sigma\right)  }%
\chi^{0\left(  \sigma\right)  }-\omega_{0\left(  \sigma\lambda\right)  }%
\chi^{0\left(  \sigma\lambda\right)  }$, because PBs among primary constraints
themselves and among primary and secondary constraints are zero.

From (\ref{eqnT49}) and (\ref{eqnT50}) and the PBs among first class
constraints we find all the functions $\alpha_{\left(  ..\right)  }^{\left(
..\right)  }\left(  x,y\right)  $ in (\ref{eqnT47a}), (\ref{eqnT47}),%

\begin{equation}
\alpha_{\left(  \alpha\beta\right)  }^{\left(  \rho\right)  }\left(
x,y\right)  =0, \label{eqnT51}%
\end{equation}

\begin{equation}
\alpha_{\left(  \gamma\right)  }^{\left(  \rho\right)  }\left(  x,y\right)
=\omega_{0(\gamma}^{~~~\rho)}\delta\left(  x-y\right)  , \label{eqnT52}%
\end{equation}

\begin{equation}
\alpha_{\left(  \gamma\right)  }^{\left(  \alpha\beta\right)  }\left(
x,y\right)  =\frac{1}{2}\left(  e_{0}^{\left(  \alpha\right)  }\delta_{\left(
\gamma\right)  }^{\left(  \beta\right)  }-e_{0}^{\left(  \beta\right)  }%
\delta_{\left(  \gamma\right)  }^{\left(  \alpha\right)  }\right)
\delta\left(  x-y\right)  , \label{eqnT54}%
\end{equation}

\begin{equation}
\alpha_{\left(  \nu\mu\right)  }^{\left(  \alpha\beta\right)  }\left(
x,y\right)  =\left(  \omega_{0~~\mu)}^{~(\alpha}\delta_{\left(  \nu\right)
}^{\left(  \beta\right)  }-\omega_{0~~\nu)}^{~(\alpha}\delta_{\left(
\mu\right)  }^{\left(  \beta\right)  }\right)  \delta\left(  x-y\right)  .
\label{eqnT53}%
\end{equation}
This completes the derivation of the generator (\ref{eqnT45}) as now%

\begin{equation}
G_{\left(  0\right)  }^{\left(  \rho\right)  }=-\chi^{0\left(  \rho\right)
}+\omega_{0(\gamma}^{~~~\rho)}\pi^{0\left(  \gamma\right)  } \label{eqnT55}%
\end{equation}
and%

\begin{equation}
G_{\left(  0\right)  }^{\left(  \alpha\beta\right)  }=-\chi^{0\left(
\alpha\beta\right)  }+\frac{1}{2}\left(  e_{0}^{\left(  \alpha\right)  }%
\delta_{\left(  \gamma\right)  }^{\left(  \beta\right)  }-e_{0}^{\left(
\beta\right)  }\delta_{\left(  \gamma\right)  }^{\left(  \alpha\right)
}\right)  \pi^{0\left(  \gamma\right)  }+\omega_{0~~\mu)}^{~(\alpha}%
\Pi^{0\left(  \beta\mu\right)  }-\omega_{0~\text{~}\mu)}^{~(\beta}%
\Pi^{0\left(  \alpha\mu\right)  }. \label{eqnT56}%
\end{equation}

Substitution of (\ref{eqnT46}), (\ref{eqnT55}), and (\ref{eqnT56}) into
(\ref{eqnT45}) gives%

\begin{align}
G  &  =\pi^{0\left(  \rho\right)  }\dot{t}_{\left(  \rho\right)  }+\left(
-\chi^{0\left(  \rho\right)  }+\omega_{0(\gamma}^{~~~\rho)}\pi^{0\left(
\gamma\right)  }\right)  t_{\left(  \rho\right)  }+\Pi^{0\left(  \alpha
\beta\right)  }\dot{r}_{\left(  \alpha\beta\right)  }+\label{eqnT56a}\\
&  \left[  -\chi^{0\left(  \alpha\beta\right)  }+\frac{1}{2}\left(
e_{0}^{\left(  \alpha\right)  }\delta_{\left(  \gamma\right)  }^{\left(
\beta\right)  }-e_{0}^{\left(  \beta\right)  }\delta_{\left(  \gamma\right)
}^{\left(  \alpha\right)  }\right)  \pi^{0\left(  \gamma\right)  }%
+\omega_{0~~\mu)}^{~(\alpha}\Pi^{0\left(  \beta\mu\right)  }-\omega
_{0~\text{~}\mu)}^{~(\beta}\Pi^{0\left(  \alpha\mu\right)  }\right]
r_{\left(  \alpha\beta\right)  }.\nonumber
\end{align}
Now using%

\begin{equation}
\delta\left(  field\right)  =\left\{  G,field\right\}  \label{eqnT57}%
\end{equation}
we can find the gauge transformations of fields.

For example, for $\delta\omega_{0\left(  \sigma\lambda\right)  }$ one finds%

\begin{equation}
\delta\omega_{0\left(  \sigma\lambda\right)  }=-\dot{r}_{\left(  \sigma
\lambda\right)  }-\left(  \omega_{0~~\lambda)}^{~(\alpha}\delta_{\left(
\sigma\right)  }^{\left(  \beta\right)  }-\omega_{0~~\sigma)}^{~(\alpha}%
\delta_{\left(  \lambda\right)  }^{\left(  \beta\right)  }\right)  r_{\left(
\alpha\beta\right)  }. \label{eqnT58}%
\end{equation}
This result is the same as Witten's \cite{Witten} for spin connections with
the \textquotedblleft temporal\textquotedblright\ external index,
$\omega_{0\left(  \sigma\lambda\right)  }$. Witten used a different notation
which is specific to $3D$, while we will present the transformations of the
fields in covariant form. Note that $\delta\omega_{0\left(  \sigma
\lambda\right)  }$ depends only on the rotational parameter. In the gauge
transformation of tetrads both parameters are present%

\begin{equation}
\delta e_{0\left(  \lambda\right)  }=-\dot{t}_{\left(  \lambda\right)
}-\omega_{0(\lambda}^{~~~\rho)}t_{\left(  \rho\right)  }-\frac{1}{2}\left(
e_{0}^{\left(  \alpha\right)  }\delta_{\left(  \lambda\right)  }^{\left(
\beta\right)  }-e_{0}^{\left(  \beta\right)  }\delta_{\left(  \lambda\right)
}^{\left(  \alpha\right)  }\right)  r_{\left(  \alpha\beta\right)  }.
\label{eqnT59}%
\end{equation}

Equation (\ref{eqnT57}) gives for $\delta e_{k\left(  \lambda\right)  }$:%

\begin{equation}
\delta e_{k\left(  \lambda\right)  }=\frac{\delta\chi^{0\left(  \rho\right)
}}{\delta\pi^{k\left(  \lambda\right)  }}t_{\left(  \rho\right)  }%
+\frac{\delta\chi^{0\left(  \alpha\beta\right)  }}{\delta\pi^{k\left(
\lambda\right)  }}r_{\left(  \alpha\beta\right)  }=-t_{\left(  \lambda\right)
,k}-\omega_{k(\lambda}^{~~~\rho)}t_{\left(  \rho\right)  }-\frac{1}{2}\left(
e_{k}^{\left(  \alpha\right)  }\delta_{\left(  \lambda\right)  }^{\left(
\beta\right)  }-e_{k}^{\left(  \beta\right)  }\delta_{\left(  \lambda\right)
}^{\left(  \alpha\right)  }\right)  r_{\left(  \alpha\beta\right)  }.
\label{eqnT60}%
\end{equation}
(Here we substituted $\pi^{k\left(  \lambda\right)  }$ in term of
$\omega_{k\left(  \alpha\beta\right)  }$ from (\ref{eqnT22}) and (\ref{eqnT23})).

We can write together (\ref{eqnT59}) and (\ref{eqnT60}) as one covariant equation%

\begin{equation}
\delta e_{\gamma\left(  \lambda\right)  }=-t_{\left(  \lambda\right)  ,\gamma
}-\omega_{\gamma(\lambda}^{~~~\rho)}t_{\left(  \rho\right)  }-\frac{1}%
{2}\left(  e_{\gamma}^{\left(  \alpha\right)  }\delta_{\left(  \lambda\right)
}^{\left(  \beta\right)  }-e_{\gamma}^{\left(  \beta\right)  }\delta_{\left(
\lambda\right)  }^{\left(  \alpha\right)  }\right)  r_{\left(  \alpha
\beta\right)  }.\label{eqnT61}%
\end{equation}
This gauge transformation, (\ref{eqnT61}), also confirms the result of Witten
\cite{Witten}, but in \cite{Witten} it was not derived.

To obtain $\delta\omega_{k\left(  \sigma\lambda\right)  }$, we need first to
find $\delta\pi^{\gamma\left(  \lambda\right)  }$. Equation (\ref{eqnT57})
gives for $\delta\pi^{\gamma\left(  \lambda\right)  }$%

\begin{equation}
\delta\pi^{0\left(  \lambda\right)  }=\frac{1}{2}\left(  \eta^{\left(
\alpha\right)  \left(  \lambda\right)  }\pi^{0\left(  \beta\right)  }%
-\eta^{\left(  \beta\right)  \left(  \lambda\right)  }\pi^{0\left(
\alpha\right)  }\right)  r_{\left(  \alpha\beta\right)  }, \label{eqnT62}%
\end{equation}

\begin{equation}
\delta\pi^{m\left(  \lambda\right)  }=\frac{1}{2}\left(  \eta^{\left(
\alpha\right)  \left(  \lambda\right)  }\pi^{m\left(  \beta\right)  }%
-\eta^{\left(  \beta\right)  \left(  \lambda\right)  }\pi^{m\left(
\alpha\right)  }\right)  r_{\left(  \alpha\beta\right)  }+eB^{m\left(
\lambda\right)  n\left(  \alpha\right)  0\left(  \beta\right)  }r_{\left(
\alpha\beta\right)  ,n}. \label{eqnT63}%
\end{equation}
Note that both equations, (\ref{eqnT62}) and (\ref{eqnT63}), can be written in
one covariant form%

\begin{equation}
\delta\pi^{\gamma\left(  \lambda\right)  }=\frac{1}{2}\left(  \eta^{\left(
\alpha\right)  \left(  \lambda\right)  }\pi^{\gamma\left(  \beta\right)
}-\eta^{\left(  \beta\right)  \left(  \lambda\right)  }\pi^{\gamma\left(
\alpha\right)  }\right)  r_{\left(  \alpha\beta\right)  }+eB^{\gamma\left(
\lambda\right)  n\left(  \alpha\right)  0\left(  \beta\right)  }r_{\left(
\alpha\beta\right)  ,n}. \label{eqnT63a}%
\end{equation}

Using (\ref{eqnT26}) and (\ref{eqnT29}), together with the transformation
properties of $e_{\gamma\left(  \lambda\right)  }$ (\ref{eqnT61}) and
$\pi^{\gamma\left(  \lambda\right)  }$ (\ref{eqnT63a}), we can obtain the
gauge transformation of $\delta\omega_{k\left(  \sigma\lambda\right)  }$:%

\begin{equation}
\delta\omega_{k\left(  \sigma\lambda\right)  }=-r_{\left(  \sigma
\lambda\right)  ,k}-\left(  \omega_{k~~\lambda)}^{~(\alpha}\delta_{\left(
\sigma\right)  }^{\left(  \beta\right)  }-\omega_{k~~\sigma)}^{~(\beta}%
\delta_{\left(  \lambda\right)  }^{\left(  \alpha\right)  }\right)  r_{\left(
\alpha\beta\right)  }. \label{eqnT64}%
\end{equation}
Now combining (\ref{eqnT58}) and (\ref{eqnT64}), we get the covariant equation
for $\delta\omega_{\gamma\left(  \sigma\lambda\right)  }$%

\begin{equation}
\delta\omega_{\gamma\left(  \sigma\lambda\right)  }=-r_{\left(  \sigma
\lambda\right)  ,\gamma}-\left(  \omega_{\gamma~~\lambda)}^{~(\alpha}%
\delta_{\left(  \sigma\right)  }^{\left(  \beta\right)  }-\omega
_{\gamma~~\sigma)}^{~(\beta}\delta_{\left(  \lambda\right)  }^{\left(
\alpha\right)  }\right)  r_{\left(  \alpha\beta\right)  }. \label{eqnT65}%
\end{equation}
The gauge transformations of field variables $e_{\gamma\left(  \lambda\right)
}$ and $\omega_{\gamma\left(  \sigma\lambda\right)  }$ have been expressed in
covariant form.

To summarize, our analysis has confirmed Witten's result: when $D=3$, we have
obtained the same gauge transformations for $e_{\gamma\left(  \lambda\right)
}$ and $\omega_{\gamma\left(  \sigma\lambda\right)  }$ as in \cite{Witten}.
From our analysis, which is based on the Dirac procedure, we derived the gauge
transformations (\ref{eqnT61}) and (\ref{eqnT65}) generated by the first class
constraints for the tetrad gravity when $D=3$. The PBs of the secondary first
class constraints form the Poincar\'{e} algebra ISO(2,1). This is not
surprising, as equivalent formulations of the same theory should produce the
same result, e.g. as Lagrangian and Hamiltonian formulations. When $D=3$, the
canonical Hamiltonian (\ref{eqnT40}) is a linear combination of the secondary
first class constraints which we have called translational and rotational, and
this is consistent with there being zero degrees of freedom (\ref{eqnT44}). We
see that the notorious \textquotedblleft
diffeomorphism\ constraint\textquotedblright\ (neither the full nor
\textquotedblleft spatial\textquotedblright\ one) does not arise in the course
of the Hamiltonian analysis of tetrad gravity in $D=3$. We would like to also
mention that Blagojevi\'{c} \cite{Blag-book} performing the Hamiltonian
analysis of the Chern-Simons action and using the Castellani procedure to find
the gauge invariance stated that \textquotedblleft the diffeomorphisms are not
found\textquotedblright\ and concluded: \textquotedblleft Thus, the
diffeomorphisms are \textit{not an independent symmetry} [Italic is of
M.B.]\textquotedblright. In the next Section we will compare the gauge
invariance found here using the Hamiltonian analysis with diffeomorphism invariance.

The last step is to check the invariance of the Lagrangian under the gauge
transformations. Actually, it is not necessary as the derivation of gauge
transformations is performed in such a way that Lagrangian should be
automatically invariant, however, we will check this for consistency. It is
not difficult to show that the transformations under rotation, the part in
(\ref{eqnT61}) proportional to $r_{\left(  \alpha\beta\right)  }$ and
(\ref{eqnT65}), give $\delta_{r}L=0$ in all dimensions ($D>2$). Note that in
derivation of rotational constraints we did not use any peculiarity of the
$D=3$ case. We also confirm in \cite{gauge} that using the Lagrangian methods
the same transformations under rotation arise in all dimensions ($D>2$) and
they leave the Lagrangian invariant. As we have shown in \cite{gauge}, the
transformations under translation are different in dimensions $D>3$. It is not
evident that the Lagrangian (\ref{eqnT1}) which has the same form in all
dimensions ($D>2$) is invariant under translational transformations that are
specific to $D=3$. From (\ref{eqnT61}) and (\ref{eqnT65}) we can see that in
the $D=3$ case they are%

\begin{equation}
\delta_{t}e_{\gamma\left(  \lambda\right)  }=-t_{\left(  \lambda\right)
,\gamma}-\omega_{\gamma(\lambda}^{~~~\rho)}t_{\left(  \rho\right)  }%
~,\qquad\delta_{t}\omega_{\gamma\left(  \sigma\lambda\right)  }%
=0.\label{eqnT66}%
\end{equation}

The proof that the EC Lagrangian (\ref{eqnT1}) is invariant under
(\ref{eqnT66}) is given in Appendix A.

\section{Gauge invariance versus diffeomorphism invariance for tetrad gravity
in D=3}

We would like to mention again that we will call the \textquotedblleft gauge
symmetry\textquotedblright\ the invariance that follows from the structure of
the first class constraints of the Hamiltonian formulation of a theory. In
particular, in the Hamiltonian analysis of the EC action in $D=3$ we found
that the gauge symmetry is translation and rotation in the tangent space. But
we know that the Lagrangian (\ref{eqnT1}) is also invariant under
diffeomorphism. Let us compare these invariances.

We will use the particular form of a diffeomorphism transformation given by
\cite{Landau, Carmeli}\footnote{In the mathematical literature, the term
diffeomorphism refers to a mapping from one manifold to another which is
differentiable, one-to-one, onto, and with a differentiable inverse.}%

\begin{equation}
\delta g_{\mu\nu}=-\xi_{\mu;\nu}-\xi_{\nu;\mu}, \label{eqnG1}%
\end{equation}
or by another equivalent form%

\begin{equation}
\delta g_{\mu\nu}=-\xi_{\mu,\nu}-\xi_{\nu,\mu}+g^{\alpha\beta}\left(
g_{\mu\beta,\nu}+g_{\nu\beta,\mu}-g_{\mu\nu,\beta}\right)  \xi_{\alpha
},\label{eqnG2}%
\end{equation}
where $\xi_{\mu}$ is the diffeomorphism parameter (which is a
\textquotedblleft world\textquotedblright\ vector) and the semicolon
\textquotedblleft$;$\textquotedblright\ signifies the covariant derivative. In
the literature on the Hamiltonian formulation of metric General Relativity the
word \textquotedblleft diffeomorphism\textquotedblright\ is often used as
being equivalent to the transformation (\ref{eqnG1}), which is similar to
gauge transformations in ordinary field theories. It is in exactly this sense
that diffeomorphism invariance was derived in the Hamiltonian analysis of the
Einstein-Hilbert action (metric gravity) when $D>2$ for the second order
\cite{KKRV, FKK} and the first order \cite{Affine} forms, without any need for
a noncovariant and/or a field dependent redefinition of the parameter
$\xi_{\mu}$.

The transformation similar to (\ref{eqnG2}) can also be derived for the tetrad
field $e_{\gamma\left(  \lambda\right)  }$. One way is to use the relation
between the metric tensor $g_{\mu\nu}$ and the tetrads $e_{\gamma\left(
\lambda\right)  }$%

\begin{equation}
g_{\mu\nu}=e_{\mu\left(  \lambda\right)  }e_{\nu}^{\left(  \lambda\right)  }.
\label{eqnG3}%
\end{equation}
From (\ref{eqnG3}) it follows that%

\begin{equation}
\delta e_{\nu\left(  \lambda\right)  }=\frac{1}{2}e_{\left(  \lambda\right)
}^{\mu}\delta g_{\mu\nu}. \label{eqnG4}%
\end{equation}
If we substitute (\ref{eqnG2}) into (\ref{eqnG4}) and use $\xi^{\rho}%
=g^{\rho\alpha}\xi_{\alpha}$, we obtain%

\begin{equation}
\delta e_{\nu\left(  \lambda\right)  }=-e_{\rho\left(  \lambda\right)  }%
\xi_{~,\nu}^{\rho}-e_{\nu\left(  \lambda\right)  ,\rho}\xi^{\rho}.
\label{eqnG5}%
\end{equation}

Another way to derive the transformation (\ref{eqnG5}) is to use the fact that
$e_{\nu\left(  \lambda\right)  }$ is a \textquotedblleft
world\textquotedblright\ vector and transforms under a general coordinate
transformations as%

\begin{equation}
e_{\left(  \lambda\right)  }^{\prime\nu}\left(  x^{\prime}\right)
=\frac{\partial x^{\prime\nu}}{\partial x^{\gamma}}e_{\left(  \lambda\right)
}^{\gamma}\left(  x\right)  . \label{eqnG5b}%
\end{equation}
For infinitesimal transformations%

\begin{equation}
x^{\mu}\rightarrow x^{\prime\mu}=x^{\mu}+\xi^{\mu}\left(  x\right)
\label{eqnG5c}%
\end{equation}
equation (\ref{eqnG5b}) can be written as%

\begin{equation}
e_{\left(  \lambda\right)  }^{\prime\mu}\left(  x^{\prime}\right)  =e_{\left(
\lambda\right)  }^{\mu}\left(  x\right)  +\xi_{,\gamma}^{\mu}e_{\left(
\lambda\right)  }^{\gamma}\left(  x\right)  +O\left(  \xi^{2}\right)  .
\label{eqnG5d}%
\end{equation}
Combining the Taylor expansion of $e_{\left(  \lambda\right)  }^{\prime\mu
}\left(  x^{\prime}\right)  $%

\begin{equation}
e_{\left(  \lambda\right)  }^{\prime\mu}\left(  x^{\prime}\right)  =e_{\left(
\lambda\right)  }^{\prime\mu}\left(  x^{\gamma}+\xi^{\gamma}\left(  x\right)
\right)  =e_{\left(  \lambda\right)  }^{\prime\mu}\left(  x\right)
+e_{\left(  \lambda\right)  ,\gamma}^{\prime\mu}\xi^{\gamma}+O\left(  \xi
^{2}\right)  \label{eqnG5e}%
\end{equation}
with (\ref{eqnG5d}) and replacing $e_{\left(  \lambda\right)  ,\gamma}%
^{\prime\mu}$ by $e_{\left(  \lambda\right)  ,\gamma}^{\mu}$, the
transformation $\delta e_{\left(  \lambda\right)  }^{\mu}\left(  x\right)  $ follows%

\begin{equation}
\delta e_{\left(  \lambda\right)  }^{\mu}\left(  x\right)  =e_{\left(
\lambda\right)  }^{\prime\mu}\left(  x\right)  -e_{\left(  \lambda\right)
}^{\mu}\left(  x\right)  =e_{\left(  \lambda\right)  }^{\gamma}\xi_{~,\gamma
}^{\mu}-e_{\left(  \lambda\right)  ,\gamma}^{\mu}\xi^{\gamma}. \label{eqnG5f}%
\end{equation}
Using $\delta\left(  \eta^{\left(  \lambda\right)  \left(  \gamma\right)
}e_{\nu\left(  \lambda\right)  }e_{\left(  \gamma\right)  }^{\mu}\right)  =0$
it is easy to obtain the transformation $\delta e_{\nu\left(  \lambda\right)
}$ given by (\ref{eqnG5}).

This perpetrated \textquotedblleft gauge\textquotedblright\ transformation of
$e_{\nu\left(  \lambda\right)  }$, (\ref{eqnG5}), can be found in many papers
on the tetrad gravity, e.g. \cite{Schwinger, DI, Carlipbook}, as well as in
Witten's paper \cite{Witten}. However, \textit{the only} gauge transformation
of $e_{\nu\left(  \lambda\right)  }$ following from the Hamiltonian
formulation is given by (\ref{eqnT61}) and is not a diffeomorphism.

As stated in \cite{Witten}, the gauge transformation (\ref{eqnT61}) and the
diffeomorphism (\ref{eqnG5}) \textquotedblleft are
equivalent\textquotedblright. However, this equivalence needs an imposition of
severe additional conditions: (i) a field-dependent redefinition of a gauge
parameter $\xi^{\beta}=e^{\beta\left(  \rho\right)  }t_{\left(  \rho\right)
}$; (ii) keeping only the translational invariance and disregarding the
rotational invariance; (iii) using the equations of motion (\textquotedblleft
on-shell\textquotedblright\ invariance). It is difficult to accept such an
\textquotedblleft equivalence\textquotedblright\ and voluntarily replace the
derived ISO(2,1) gauge symmetry of tetrad gravity by diffeomorphism plus
Lorentz invariance or, even worse, with only a \textquotedblleft
spatial\textquotedblright\ diffeomorphism, as is often presented in the
literature. As we have already shown, a gauge invariance of a theory can be
found exactly if one follows the Dirac procedure in which one casts the theory
into a Hamiltonian form, finds all constraints, the PBs among them and
classifies them as first class or second class, derives the gauge generator
from the first class constraints, and finally uses this gauge generator to
find gauge transformations of variables in the theory. Using this procedure,
we have derived the gauge transformations of tetrads $e_{\gamma\left(
\lambda\right)  }$ (\ref{eqnT61}) and spin connections $\omega_{\rho\left(
\alpha\beta\right)  }$ (\ref{eqnT65}). Moreover, the algebra of secondary
first class constraints gives unambiguously the Poincar\'{e} algebra ISO(2,1),
(\ref{qnT41})-(\ref{eqnT43}), not an algebra of diffeomorphism and Lorentz
rotations. We have to conclude that the gauge invariance of the tetrad gravity
in three dimensions is a Poincar\'{e} symmetry. Diffeomorphism (\ref{eqnG5})
is the\ symmetry of the Einstein-Cartan action, but it is NOT A GAUGE SYMMETRY
derived from the first class constraints in the Hamiltonian formulation of
tetrad gravity \cite{gauge}.

It is not surprising that metric and tetrad gravity theories have different
gauge symmetries as they are not equivalent. Einstein in his article on tetrad
($n$-bein) gravity wrote \cite{Einstein1928}: \textquotedblleft The $n$-bein
field is determined by $n^{2}$ functions $h_{a}^{\mu}$ [tetrads $e_{\left(
\alpha\right)  }^{\mu}$, in our notation], whereas the Riemannian metric is
determined by $\frac{n\left(  n+1\right)  }{2}$ quantities. According to (3)
[$g_{\mu\nu}=h_{\mu a}h_{\nu}^{a}$], the metric is determined by the $n$-bein
field but not vice versa\textquotedblright. So, the attempt to deduce a gauge
transformation of $e_{\gamma\left(  \lambda\right)  }$ from the diffeomorphism
invariance of $g_{\mu\nu}$ is the wrong way. But it should be possible to
deduce a transformation of $g_{\mu\nu}$ from that of $e_{\gamma\left(
\lambda\right)  }$ (the \textquotedblleft vice versa\textquotedblright\ of Einstein).

To compare the results of (\ref{eqnT61}) and (\ref{eqnG5}), without forcing an
equivalence by imposing the restrictions (i)-(iii) mentioned after
(\ref{eqnG5f}), we rewrite (\ref{eqnT61}) in a slightly different form. First
of all, from the equation of motion $\frac{\delta L}{\delta\omega_{\mu\left(
\alpha\beta\right)  }}=0$ it follows%

\begin{equation}
B^{\varepsilon\left(  \alpha\right)  \mu\left(  \lambda\right)  \sigma\left(
\rho\right)  }e_{\varepsilon\left(  \alpha\right)  ,\mu}-A^{\sigma\left(
\lambda\right)  \nu\left(  \beta\right)  }\omega_{\nu~~\beta)}^{~(\rho
}+A^{\sigma\left(  \rho\right)  \nu\left(  \beta\right)  }\omega_{\nu~~\beta
)}^{~(\lambda}=0. \label{eqn-eom}%
\end{equation}
Solving (\ref{eqn-eom}) for $\omega_{\nu~~\beta)}^{~~(\rho}$, we can express
it in terms of $e_{\nu\left(  \lambda\right)  }$ and its derivatives%

\begin{equation}
\omega_{\sigma}^{~\left(  \alpha\beta\right)  }=\frac{1}{2}e_{\sigma\left(
\lambda\right)  }e_{\varepsilon\left(  \rho\right)  ,\mu}\left(  \eta^{\left(
\rho\right)  \left(  \beta\right)  }A^{\varepsilon\left(  \alpha\right)
\mu\left(  \lambda\right)  }+\eta^{\left(  \rho\right)  \left(  \alpha\right)
}A^{\varepsilon\left(  \lambda\right)  \mu\left(  \beta\right)  }%
-\eta^{\left(  \rho\right)  \left(  \lambda\right)  }A^{\varepsilon\left(
\beta\right)  \mu\left(  \alpha\right)  }\right)  , \label{eqnG5a}%
\end{equation}
or in more familiar form%

\begin{equation}
\omega_{\sigma}^{~\left(  \alpha\beta\right)  }=\frac{1}{2}\left[
e^{\varepsilon\left(  \alpha\right)  }\left(  e_{\varepsilon,~\sigma}^{\left(
\beta\right)  }-e_{\sigma,~\varepsilon}^{\left(  \beta\right)  }\right)
-e^{\varepsilon\left(  \beta\right)  }\left(  e_{\varepsilon,~\sigma}^{\left(
\alpha\right)  }-e_{\sigma,~\varepsilon}^{\left(  \alpha\right)  }\right)
-e_{\sigma}^{\left(  \rho\right)  }e^{\varepsilon\left(  \beta\right)  }%
e^{\mu\left(  \alpha\right)  }\left(  e_{\varepsilon\left(  \rho\right)  ,\mu
}-e_{\mu\left(  \rho\right)  ,\varepsilon}\right)  \right]  . \label{eqnG5-1}%
\end{equation}

Substitution of (\ref{eqnG5a}) into (\ref{eqnT61}) gives%

\[
\delta e_{\gamma\left(  \lambda\right)  }=-t_{\left(  \lambda\right)  ,\gamma
}-\frac{1}{2}\left(  e_{\gamma\left(  \lambda\right)  ,\varepsilon
}-e_{\varepsilon\left(  \lambda\right)  ,\gamma}\right)  e^{\varepsilon\left(
\rho\right)  }t_{\left(  \rho\right)  }-\frac{1}{2}e_{\gamma}^{\left(
\alpha\right)  }e_{\left(  \lambda\right)  }^{\varepsilon}\left(
e_{\varepsilon\left(  \alpha\right)  ,\mu}-e_{\mu\left(  \alpha\right)
,\varepsilon}\right)  e^{\mu\left(  \rho\right)  }t_{\left(  \rho\right)  }%
\]

\begin{equation}
-\frac{1}{2}\left(  e_{\varepsilon\left(  \mu\right)  ,\gamma}-e_{\gamma
\left(  \mu\right)  ,\varepsilon}\right)  e_{\left(  \lambda\right)
}^{\varepsilon}t^{\left(  \mu\right)  }-\frac{1}{2}\left(  e_{\gamma}^{\left(
\alpha\right)  }\delta_{\left(  \lambda\right)  }^{\left(  \beta\right)
}-e_{\gamma}^{\left(  \beta\right)  }\delta_{\left(  \lambda\right)
}^{\left(  \alpha\right)  }\right)  r_{\left(  \alpha\beta\right)  }.
\label{eqnG6a}%
\end{equation}
From (\ref{eqnG3}) we can find that%

\begin{equation}
\delta g_{\mu\nu}=e_{\mu}^{\left(  \lambda\right)  }\delta e_{\nu\left(
\lambda\right)  }+e_{\nu}^{\left(  \lambda\right)  }\delta e_{\mu\left(
\lambda\right)  }. \label{eqnG7}%
\end{equation}
Now from $\delta e_{\mu\left(  \lambda\right)  }$, (\ref{eqnG6a}), and from
(\ref{eqnG7}) we can obtain $\delta g_{\mu\nu}$%

\begin{equation}
\delta g_{\mu\nu}=-\left(  e_{\mu}^{\left(  \rho\right)  }t_{\left(
\rho\right)  }\right)  _{,\nu}-\left(  e_{\nu}^{\left(  \rho\right)
}t_{\left(  \rho\right)  }\right)  _{,\mu}+\left(  g_{\mu\beta,\nu}%
+g_{\nu\beta,\mu}-g_{\mu\nu,\beta}\right)  e^{\beta\left(  \rho\right)
}t_{\left(  \rho\right)  }\label{eqnG8}%
\end{equation}
that after the redefinition $\xi^{\beta}=e^{\beta\left(  \rho\right)
}t_{\left(  \rho\right)  }$ leads to (\ref{eqnG2}). Note, that contributions
with the rotational parameter $r_{\left(  \alpha\beta\right)  }$ from
(\ref{eqnG6a}) completely cancel out in (\ref{eqnG8}), as well as some terms
proportional to $t_{\left(  \rho\right)  }$, without imposing any conditions.
We do not need the additional restrictions (ii)-(iii); only field-dependent
redefinition of parameters (i) is needed. Thus it is possible to obtain the
diffeomorphism invariance of $g_{\mu\nu}$ from the Poincar\'{e} symmetry of
$e_{\gamma\left(  \lambda\right)  }$, but not vice versa. A field-dependent
redefinition of the gauge parameter is necessary but this is a consequence of
the non-equivalence of metric and tetrad gravities. A similar redefinition had
to be introduced when we considered the two dimensional metric and tetrad
gravities in \cite{2DGKK}, despite that the gauge symmetry of two dimensional
gravity (both metric and tetrad) is very different from higher dimensions.

\section{Conclusion}

In his book \cite{Diracbook} Dirac wrote \textquotedblleft\textit{I
}[Dirac]\textit{ feel that there will always be something missing from them
}[non-Hamiltonian methods]\textit{ which we can only get by working from a
Hamiltonian}\textquotedblright. We feel that the Hamiltonian method not only
allows one to find something that can be missed when using other methods but
also protects us from \textquotedblleft finding\textquotedblright\ something
that might be attributed to a theory but really not there. In particular, the
gauge invariance of a theory should follow from the Dirac procedure and any
guess (even an intelligent one) has to be supported by calculations.

The results reported in this paper confirm Dirac's old conjecture
\cite{Diracbook}. The Hamiltonian formulation of the tetrad gravity considered
here using the Dirac procedure leads to self-consistent and unambiguous
results. In particular, this approach gives a unique answer to the question of
what is the true gauge invariance of the tetrad gravity and eliminates any
possibility of being able to \textquotedblleft choose\textquotedblright\ a
gauge invariance based on either a belief, desire or common wisdom, because
the gauge invariance should be derivable from the unique constraint structure
of this (or any) theory. This constraint structure can be modified only by a
non-canonical change of variables that immediately destroys any connection
with an original theory and, at best, can be considered as some model not
related to the tetrad gravity (see the discussion on non-canonical change of
variables for tetrad gravity in Section 5 of \cite{Affine}).

After Hamiltonian reduction, solving the second class constraints and
eliminating non-physical (redundant) variables, the remaining first class
constraints form a Lie algebra. This is exactly what one can expect for the
local field theory and this is precisely what one needs to quantize it. The
results presented in this paper were mainly obtained for $D=3$ case which has
been treated completely. The Hamiltonian analysis of the EC action in higher
dimensions is in progress \cite{Report, Darboux} and further developments will
be reported elsewhere.

However, even results of this paper provide enough information to form some
conclusions about the Hamiltonian formulation of tetrad gravity in higher
dimensions, based on the following reasoning:

- the Lagrangian of the first order formulation of the tetrad gravity
(\ref{eqnT2}) with tetrads and spin connections treated as independent fields
gives the equivalent equations of motion as its second order counterpart, and
this can be demonstrated in any dimensions $D>2$ in covariant form and without
any recourse to a particular dimension and does not show any peculiarities in
$D=3$;

- the primary first class constraints are a part of the generator and
unambiguously define the gauge parameters in (\ref{eqnT45}); in any dimension,
because of the antisymmetry of $B^{\gamma\left(  \rho\right)  \mu\left(
\alpha\right)  \nu\left(  \beta\right)  }$, there is a translational
constraint $\pi^{0\left(  \rho\right)  }$ and the corresponding gauge
parameter cannot have a \textquotedblleft world\textquotedblright\ index that
we would need if diffeomorphism were a gauge invariance derivable from the
first class constraints;

- the explicit form of the rotational constraint $\chi^{0\left(  \alpha
\beta\right)  }$ (\ref{eqnT36}) is not peculiar to being $D=3$ and written in
covariant form, that is why it remains unchanged in dimensions $D>3$
\cite{Report, Darboux};

- the PBs among the rotational constraints (\ref{eqnT43}) were also calculated
without using any specific property of $D=3$;

- the final form of gauge transformations of the tetrad gravity when $D=3$ can
be cast into a covariant dimension-independent form (\ref{eqnT61},
\ref{eqnT65});

- the secondary translational constraint $\chi^{0\left(  \rho\right)  }$ has
specific to $D=3$ form (the second term in (\ref{eqnT35})) but it does not
mean that it will be modified in higher dimensions in such a way that it will
destroy the translational invariance in the internal space.

Based on the above arguments, we can conclude that diffeomorphism invariance
is not \textit{a gauge symmetry} derived from the first class constraints of
the tetrad gravity, neither in $D=3$ nor in higher dimensions, and the
translational and rotational invariance are expected in all dimensions ($D>2$).

These conclusions for $D>3$ can be called conjectures. The only way to prove
or disprove them is to apply the Dirac procedure and explicitly find the
Hamiltonian, eliminate all second class constraints and use all remaining
first class constraints to build the generators that will produce the true
gauge invariance of the tetrad gravity in higher dimensions.\footnote{Of
course, a non-canonical change of variables has to be excluded at any step of
calculations, and any manipulation that destroys the equivalence with the
original theory are not permissible.} Some preliminary results are reported in
\cite{Report, gauge, Darboux}.

\vspace{1cm}

\textbf{Acknowledgements}

We would like to thank D.G.C. McKeon for helpful discussions during the
preparation of the paper and reading the manuscript.

\appendix

\section{ $ABC$ properties and translational invariance of the EC Lagrangian}

Here we collect properties of the $ABC$ functions that are useful in the
Hamiltonian analysis \cite{Report, Darboux} and also in the Lagrangian
approach \cite{gauge} to the Einstein-Cartan action.

These functions are generated by consecutive variation of the $n$-bein density
$ee^{\mu\left(  \alpha\right)  }$:%

\begin{equation}
\frac{\delta}{\delta e_{\nu\left(  \beta\right)  }}\left(  ee^{\mu\left(
\alpha\right)  }\right)  =e\left(  e^{\mu\left(  \alpha\right)  }e^{\nu\left(
\beta\right)  }-e^{\mu\left(  \beta\right)  }e^{\nu\left(  \alpha\right)
}\right)  =eA^{\mu\left(  \alpha\right)  \nu\left(  \beta\right)  },
\label{eqn40}%
\end{equation}

\begin{equation}
\frac{\delta}{\delta e_{\lambda\left(  \gamma\right)  }}\left(  eA^{\mu\left(
\alpha\right)  \nu\left(  \beta\right)  }\right)  =eB^{\lambda\left(
\gamma\right)  \mu\left(  \alpha\right)  \nu\left(  \beta\right)  },\text{ \ }
\label{eqn41}%
\end{equation}

\begin{equation}
\frac{\delta}{\delta e_{\tau\left(  \sigma\right)  }}\left(  eB^{\lambda
\left(  \gamma\right)  \mu\left(  \alpha\right)  \nu\left(  \beta\right)
}\right)  =eC^{\tau\left(  \sigma\right)  \lambda\left(  \gamma\right)
\mu\left(  \alpha\right)  \nu\left(  \beta\right)  },\quad... \label{eqn41a}%
\end{equation}

The first important property of these density functions is their total
antisymmetry: interchange of two indices of the same nature (internal or
external), e.g.%

\begin{equation}
A^{\nu\left(  \beta\right)  \mu\left(  \alpha\right)  }=-A^{\nu\left(
\alpha\right)  \mu\left(  \beta\right)  }=-A^{\mu\left(  \beta\right)
\nu\left(  \alpha\right)  } \label{eqn42}%
\end{equation}
with the same being valid for $B$, $C$, etc. In calculations presented here,
nothing is needed beyond $C$.

The second important property is their expansion using an external index%

\begin{equation}
B^{\tau\left(  \rho\right)  \mu\left(  \alpha\right)  \nu\left(  \beta\right)
}=e^{\tau\left(  \rho\right)  }A^{\mu\left(  \alpha\right)  \nu\left(
\beta\right)  }+e^{\tau\left(  \alpha\right)  }A^{\mu\left(  \beta\right)
\nu\left(  \rho\right)  }+e^{\tau\left(  \beta\right)  }A^{\mu\left(
\rho\right)  \nu\left(  \alpha\right)  }, \label{eqn43}%
\end{equation}

\begin{equation}
C^{\tau\left(  \rho\right)  \lambda\left(  \sigma\right)  \mu\left(
\alpha\right)  \nu\left(  \beta\right)  }=e^{\tau\left(  \rho\right)
}B^{\lambda\left(  \sigma\right)  \mu\left(  \alpha\right)  \nu\left(
\beta\right)  }-e^{\tau\left(  \sigma\right)  }B^{\lambda\left(
\alpha\right)  \mu\left(  \beta\right)  \nu\left(  \rho\right)  }%
+e^{\tau\left(  \alpha\right)  }B^{\lambda\left(  \beta\right)  \mu\left(
\rho\right)  \nu\left(  \sigma\right)  }-e^{\tau\left(  \beta\right)
}B^{\lambda\left(  \rho\right)  \mu\left(  \sigma\right)  \nu\left(
\alpha\right)  } \label{eqn44}%
\end{equation}
or an internal index%

\begin{equation}
B^{\tau\left(  \rho\right)  \mu\left(  \alpha\right)  \nu\left(  \beta\right)
}=e^{\tau\left(  \rho\right)  }A^{\mu\left(  \alpha\right)  \nu\left(
\beta\right)  }+e^{\mu\left(  \rho\right)  }A^{\nu\left(  \alpha\right)
\tau\left(  \beta\right)  }+e^{\nu\left(  \rho\right)  }A^{\tau\left(
\alpha\right)  \mu\left(  \beta\right)  }, \label{eqn45}%
\end{equation}

\begin{equation}
C^{\tau\left(  \rho\right)  \lambda\left(  \sigma\right)  \mu\left(
\alpha\right)  \nu\left(  \beta\right)  }=e^{\tau\left(  \rho\right)
}B^{\lambda\left(  \sigma\right)  \mu\left(  \alpha\right)  \nu\left(
\beta\right)  }-e^{\lambda\left(  \rho\right)  }B^{\mu\left(  \sigma\right)
\nu\left(  \alpha\right)  \tau\left(  \beta\right)  }+e^{\mu\left(
\rho\right)  }B^{\nu\left(  \sigma\right)  \tau\left(  \alpha\right)
\lambda\left(  \beta\right)  }-e^{\nu\left(  \rho\right)  }B^{\tau\left(
\sigma\right)  \lambda\left(  \alpha\right)  \mu\left(  \beta\right)  }.
\label{eqn46}%
\end{equation}

The third property involves their derivatives%

\begin{equation}
\left(  eA^{\nu\left(  \beta\right)  \mu\left(  \alpha\right)  }\right)
,_{\sigma}=\frac{\delta}{\delta e_{\lambda\left(  \gamma\right)  }}\left(
eA^{\nu\left(  \beta\right)  \mu\left(  \alpha\right)  }\right)
e_{\lambda\left(  \gamma\right)  ,\sigma}=eB^{\lambda\left(  \gamma\right)
\nu\left(  \beta\right)  \mu\left(  \alpha\right)  }e_{\lambda\left(
\gamma\right)  ,\sigma}\ ,\label{eqn47}%
\end{equation}

\begin{equation}
\left(  eB^{\tau\left(  \rho\right)  \nu\left(  \beta\right)  \mu\left(
\alpha\right)  }\right)  ,_{\sigma}=\frac{\delta}{\delta e_{\lambda\left(
\gamma\right)  }}\left(  eB^{\tau\left(  \rho\right)  \nu\left(  \beta\right)
\mu\left(  \alpha\right)  }\right)  e_{\lambda\left(  \gamma\right)  ,\sigma
}=eC^{\tau\left(  \rho\right)  \lambda\left(  \gamma\right)  \nu\left(
\beta\right)  \mu\left(  \alpha\right)  }e_{\tau\left(  \rho\right)  ,\sigma
}\ .\label{eqn46a}%
\end{equation}

Upon using the antisymmetry of $B$ both in the external and internal indices
and the antisymmetry of $\omega$ in its internal indices leads to%

\begin{equation}
B^{\tau\left(  \rho\right)  \mu\left(  \alpha\right)  \nu\left(  \beta\right)
}\omega_{\mu\left(  \alpha\gamma\right)  }\omega_{\nu~~\sigma)}^{~(\gamma
}\omega_{\tau~~\beta)}^{~(\sigma}=0\label{eqn50b}%
\end{equation}
and%

\begin{equation}
eB^{\tau\left(  \rho\right)  \mu\left(  \alpha\right)  \nu\left(
\beta\right)  }\omega_{\nu\left(  \alpha\beta\right)  ,\mu\tau}%
=0.\label{eqn50c}%
\end{equation}

The above properties considerably simplify calculations. The list of $ABC$
properties can be extended, but for our purpose the above relations are adequate.

Let us check the invariance of the Einstein-Cartan Lagrangian\ (\ref{eqnT1})
under the translational transformations in $D=3$ (\ref{eqnT66}):%

\begin{align*}
\delta_{t}L  & =\delta_{t}\left[  -eA^{\mu\left(  \alpha\right)  \nu\left(
\beta\right)  }\left(  \omega_{\nu\left(  \alpha\beta\right)  ,\mu}%
+\omega_{\mu\left(  \alpha\sigma\right)  }\omega_{\nu~~\beta)}^{~(\sigma
}\right)  \right]  =-\delta_{t}\left(  eA^{\mu\left(  \alpha\right)
\nu\left(  \beta\right)  }\right)  \left(  \omega_{\nu\left(  \alpha
\beta\right)  ,\mu}+\omega_{\mu\left(  \alpha\sigma\right)  }\omega
_{\nu~~\beta)}^{~(\sigma}\right)  \\
& =-eB^{\gamma\left(  \lambda\right)  \mu\left(  \alpha\right)  \nu\left(
\beta\right)  }\delta_{t}e_{\gamma\left(  \lambda\right)  }\left(  \omega
_{\nu\left(  \alpha\beta\right)  ,\mu}+\omega_{\mu\left(  \alpha\sigma\right)
}\omega_{\nu~~\beta)}^{~(\sigma}\right)  \\
& =-eB^{\gamma\left(  \lambda\right)  \mu\left(  \alpha\right)  \nu\left(
\beta\right)  }\left(  -t_{\left(  \lambda\right)  ,\gamma}-\omega
_{\gamma(\lambda}^{~~~\rho)}t_{\left(  \rho\right)  }\right)  \left(
\omega_{\nu\left(  \alpha\beta\right)  ,\mu}+\omega_{\mu\left(  \alpha
\sigma\right)  }\omega_{\nu~~\beta)}^{~(\sigma}\right)  .
\end{align*}
After extracting the total derivative it leads to the following%

\begin{align}
\delta_{t}L  & =-\left[  eB^{\gamma\left(  \lambda\right)  \mu\left(
\alpha\right)  \nu\left(  \beta\right)  }t_{\left(  \lambda\right)  }\left(
\omega_{\nu\left(  \alpha\beta\right)  ,\mu}+\omega_{\mu\left(  \alpha
\sigma\right)  }\omega_{\nu~~\beta)}^{~(\sigma}\right)  \right]  _{,\gamma
}\label{eqn53a}\\
& +\left(  eB^{\gamma\left(  \lambda\right)  \mu\left(  \alpha\right)
\nu\left(  \beta\right)  }\right)  _{,\gamma}t_{\left(  \lambda\right)
}\left(  \omega_{\nu\left(  \alpha\beta\right)  ,\mu}+\omega_{\mu\left(
\alpha\sigma\right)  }\omega_{\nu~~\beta)}^{~(\sigma}\right)  \nonumber\\
& +eB^{\gamma\left(  \lambda\right)  \mu\left(  \alpha\right)  \nu\left(
\beta\right)  }t_{\left(  \lambda\right)  }\omega_{\nu\left(  \alpha
\beta\right)  ,\mu\gamma}+eB^{\gamma\left(  \lambda\right)  \mu\left(
\alpha\right)  \nu\left(  \beta\right)  }t_{\left(  \lambda\right)  }\left(
\omega_{\mu\left(  \alpha\sigma\right)  ,\gamma}\omega_{\nu~~\beta)}%
^{~(\sigma}+\omega_{\mu\left(  \alpha\sigma\right)  }\omega_{\nu
~~\beta),\gamma}^{~(\sigma}\right)  \nonumber\\
& -eB^{\gamma\left(  \lambda\right)  \mu\left(  \alpha\right)  \nu\left(
\beta\right)  }t_{\left(  \rho\right)  }\omega_{\gamma(\lambda}^{~~~\rho
)}\omega_{\nu\left(  \alpha\beta\right)  ,\mu}-eB^{\gamma\left(
\lambda\right)  \mu\left(  \alpha\right)  \nu\left(  \beta\right)  }t_{\left(
\rho\right)  }\omega_{\gamma(\lambda}^{~~~\rho)}\omega_{\mu\left(
\alpha\sigma\right)  }\omega_{\nu~~\beta)}^{~(\sigma}.\nonumber
\end{align}

The first term in (\ref{eqn53a}) is the total derivative, the second term is
zero in $D=3$. This is because the derivative $\left(  eB^{\gamma\left(
\lambda\right)  \mu\left(  \alpha\right)  \nu\left(  \beta\right)  }\right)
_{,\gamma}$ is proportional to $C$ which is antisymmetric in four internal and
four external indices, but in $D=3$ there are only three distinct indices:
$0,1,2$. The third term in (\ref{eqn53a}) is also zero because of
(\ref{eqn50c}). It is not evident that the fourth and fifth terms cancel. From
the antisymmetry of $B$ it follows that $\lambda\neq\alpha\neq\beta$ in these
terms; in addition, from the antisymmetry of $\omega$ in internal indices and
from the fact that in $D=3$ there are only three distinct indices we can
conclude that in the fifth term $\rho$ should be equal either $\alpha$ or
$\beta$ (note that this is a peculiarity of $D=3$ only). If we consider both
cases, $\rho=\alpha$ and $\rho=\beta$, and relabel dummy indices in the fifth
term, we find that the fifth and fourth terms cancel. The last, sixth, term is
not exactly in the form of (\ref{eqn50b}). However, if we use again
antisymmetry of $B$ and $\omega$ and peculiarity of $D=3$ (only three distinct
indices), then we have also only two cases: either $\rho=\alpha$ or
$\rho=\beta$. In both cases we can apply (\ref{eqn50b}) for the sixth term.
Finally, under translational transformations of (\ref{eqnT66}) in $D=3$ we have%

\[
\delta_{t}L=-\left[  eB^{\gamma\left(  \lambda\right)  \mu\left(
\alpha\right)  \nu\left(  \beta\right)  }t_{\left(  \lambda\right)  }\left(
\omega_{\nu\left(  \alpha\beta\right)  ,\mu}+\omega_{\mu\left(  \alpha
\sigma\right)  }\omega_{\nu~~\beta)}^{~(\sigma}\right)  \right]  _{,\gamma}%
\]
or, using definition of $B$ (\ref{eqn43}) the Ricci tensor $R_{\mu\nu\left(
\alpha\beta\right)  }=\omega_{\nu\left(  \alpha\beta\right)  ,\mu}-\omega
_{\mu\left(  \alpha\beta\right)  ,\nu}+\omega_{\mu\left(  \alpha\sigma\right)
}\omega_{\nu~~\beta)}^{~(\sigma}-\omega_{\nu\left(  \alpha\sigma\right)
}\omega_{\mu~~\beta)}^{~(\sigma}$,%

\[
\delta_{t}L=-\left[  e\left(  e^{\gamma\left(  \lambda\right)  }e^{\mu\left(
\alpha\right)  }e^{\nu\left(  \beta\right)  }+e^{\gamma\left(  \alpha\right)
}e^{\mu\left(  \beta\right)  }e^{\nu\left(  \lambda\right)  }+e^{\gamma\left(
\beta\right)  }e^{\mu\left(  \lambda\right)  }e^{\nu\left(  \alpha\right)
}\right)  t_{\left(  \lambda\right)  }R_{\mu\nu\left(  \alpha\beta\right)
}\right]  _{,\gamma}.
\]

\end{document}